\documentclass[11pt]{article}
\usepackage{graphics}
\usepackage{amscd}
\usepackage{psfrag}
\usepackage{amsmath}

\newcommand{\be}{\begin{equation}}
\newcommand{\ee}{\end{equation}}
\newcommand{\bea}{\begin{eqnarray}}
\newcommand{\eea}{\end{eqnarray}}
\newcommand{\bay}{\begin{array}}
\newcommand{\eay}{\end{array}}
\newcommand{\pal}{\partial}
\newcommand{\ba}{\begin{eqnarray}}
\newcommand{\ea}{\end{eqnarray}}
\newcommand{\bi}{\begin{itemize}}
\newcommand{\ei}{\end{itemize}}
\newcommand{\non}{\nonumber}
\newcommand{\pho}{\phantom{-}}

\newcommand{\pbf}{{\bf p}}
\newcommand{\qbf}{{\bf q}}
\newcommand{\xbf}{{\bf x}}
\newcommand{\ybf}{{\bf y}}
\newcommand{\zerobf}{{\bf 0}}

\title{
\rightline{\small HU-EP-08/30}\vspace{-2mm}
\rightline{\small SFB/CPP-08-61}
Automatic generation of Feynman rules
\\
in the Schr\"odinger functional
}

\author{Shinji Takeda\\
Humboldt Universit\"at zu Berlin,\\
Newtonstr.~15, 12489 Berlin, Germany.}

\begin{document}
\maketitle
\begin{abstract}
We provide an algorithm to generate
vertices for the Schr\"odinger functional
with an abelian background gauge field.
The background field has a non-trivial color structure,
therefore we mainly focus on a manipulation of the color matrix part.
We propose how to implement the algorithm especially in python code.
By using python outputs produced by the code,
we also show how to write a numerical expression of vertices
in the time-momentum as well as the coordinate space
into a Feynman diagram calculation code.
As examples of the applications of the algorithm,
we provide some one-loop results,
ratios of the $\Lambda$ parameters between
the plaquette gauge action and the improved gauge actions
composed from six-link loops (rectangular, chair and parallelogram),
the determination of the O($a$) boundary counter term to this order,
and the perturbative
cutoff effects of the step scaling function of the
Schr\"odinger functional
coupling constant.
\end{abstract}

\section{Introduction}
\label{sec:introduction}
It is well known that
vertices in lattice gauge theory
are quite complicated
especially for the pure gauge sector.
This is because, on the lattice,
while gauge invariance is preserved exactly at a finite lattice spacing,
the lattice regularization itself breaks the Lorentz symmetry explicitly.
This is the main difficulty of the lattice perturbation theory.
To reduce the risk of errors and to alleviate the  tedious task
of deriving the vertices,
it is desirable to have an automatic method.
A first attempt was made by L\"uscher and Weisz about twenty
yeas ago~\cite{Luscher:1985wf}.
They worked in momentum space and
performed some calculations by using their algorithm which
is restricted to closed gauge loops but sufficient
for pure gauge theory.
About two decades later a new algorithm, which we call bottom up algorithm,
was proposed by Hart et.~al.~\cite{Hart:2004bd}.
A crucial point in this generalization  is that
it can deal with any parallel transporter,
not only with closed loops.
This algorithm also allows us to extend to fermion actions
or even smeared HQET actions.
In fact, it is applied to the NRQCD~\cite{Hart:2006ij}.
A relevant assumption of this algorithm has been translation invariance.
Needless to say their algorithm also assumes that
the background field is set to a trivial one.

Our main concern in this paper is
to extend the bottom up algorithm to the Schr\"odinger functional
(SF) \cite{Luscher:1993gh}, where
translation invariance for the time direction is broken,
and there is an abelian color background gauge field.
An essential question here is
how to deal with the non-trivial color matrix
which involves the complicated background field.
Actually this issue can be solved by noting
two important properties of the background field,
that is, abelian and a linear time dependence.
Although the latter property can be relaxed,
exploiting this is greatly helpful to gain speed up of the calculations.
Our main purpose in this paper is an exposition of the solution.
The procedure to generate vertices consists of two parts,
first, generating lists which contain information of
the non-zero elements of the vertices, and second,
writing down a numerical expression of the vertex
by making use of the lists as an input.
In this paper, we will describe how to write down
the vertex in the Feynman diagram calculation code
not only for the time-momentum space but also the position space.
In the two-loop calculation of the SF coupling~\cite{Narayanan:1995ex},
it was emphasized that
the position space calculation has
the advantage of substantially reducing the computational efforts
over the time-momentum space.
This is the reason why we consider the position space vertex here.
As it is the case for the original bottom up algorithm,
our generalization to the SF
is in principle applicable to both fermion and gauge actions.

The rest of this paper is organized as follows.
First we review the bottom up algorithm in the next section.
In section \ref{sec:SF},
after a brief reminder of the SF,
we provide a nice representation of link variables and
key equations derived from a special property of the
background field,
and then we show the algorithm to generate
lists for the SF.
This section is the main part of the paper.
Then, an explanation of how to write down the vertices
to a program which calculates Feynman diagrams
(in particular the Big Mac diagram~\cite{Narayanan:1995ex})
is given in section \ref{sec:WritingVertex}.
As mentioned before
we provide it in both the time-momentum and the position space.
As an application, in section \ref{sec:application},
we give a one-loop calculation
of the SF coupling for various gauge actions including six-link loops.
Our conclusion is given in the last section, 
together with comments on possible future work.
Some technical details are given in appendices, like
how to partially symmetrize the vertex
for the gauge action in section \ref{sec:symmetrization},
further reduction of the number of lists in \ref{sec:reduction},
and how to obtain the $\eta$ derivative of the vertex in
\ref{sec:etaderivative} respectively.

\section{Bottom up Algorithm}
\label{sec:bottomup}
In this section, we summarize the 
position space version of the bottom up algorithm on
the usual translation invariant lattice (non SF).
The explanation here is mainly based on~\cite{Hart:2006ij}
apart from the position space.
We set up the theory on a hyper-cubic euclidean lattice $\Lambda$
with spacing $a$ and size $L^3\times T$.
We assume periodic boundary conditions for all directions.
A coordinate $x \in \Lambda$ has components
$x_\mu$ with $\mu=1,2,3,4$.
In the following we denote 
the link variable (still without background field) by
\be
U(x,\mu)=\exp (a g_0 q_{\mu}(x)),
\ee
and
the anti-hermitian gauge
fluctuation field by
$q_{\mu}(x)=\sum_a q_{\mu}^a(x)T^a$, where
$T^a$ are anti-hermitian and are elements of the Lie algebra of $SU(3)$.
They are explicitly given in Appendix \ref{sec:liealgebra}.

A first important point for the automatic operation
is how to represent the vertices in a program.
As an example, we consider a parallel transporter
$P({\cal L},x_{\rm s},x_{\rm e})$ along a curve
${\cal L}$ on the lattice. We denote with
$x_{\rm s}$ and $x_{\rm e}$ the start and end point
of the parallel transporter.
Taylor expansion of the parallel transporter is written as
\bea
P({\cal L},x_{\rm s},x_{\rm e})
&=&
1+
\sum_{r=1}^{\infty}
\frac{(ag_0)^r}{r!}
\sum_{\alpha_1,\cdots,\alpha_r}
\sum_{a_1,\cdots,a_r}
V^{a_1 \cdots a_r}_{\alpha_1 \cdots \alpha_r}
\prod_{j=1}^r q^{a_j}_{\alpha_j},
\label{eqn:P}
\\
V^{a_1 \cdots a_r}_{\alpha_1 \cdots \alpha_r}
&=&
{\cal C}^{a_1\cdots a_r}
Y_{\alpha_1 \cdots \alpha_r},
\label{eqn:VcY}
\eea
where $\alpha=(\mu,x)$ is a combined index labelling links.
The (unsymmetrized) vertex
$V^{a_1 \cdots a_r}_{\alpha_1 \cdots \alpha_r}$,
has been factorized\footnote{
For the SF, however, this nice structure is rendered more complicated
by the presence of the background field as we will see in the next section.
}
into the color factor 
\be
{\cal C}^{a_1\cdots a_r}=T^{a_1}\cdots T^{a_r},
\ee
and the reduced vertex, $Y_{\alpha_1 \cdots \alpha_r}$.
The color factor is independent of the shape of the parallel transporter,
therefore we consider only the reduced vertex.

For a usual parallel transporter,
almost all elements of the corresponding reduced vertex are zero.
Therefore usually only non-zero elements are stored in a memory of a
machine. This is easily realized as follows.
First one labels the non-zero elements
of the reduced vertex with number $i$ properly ($i=1,...,N_r$), and then
one stores its indices, $\alpha_{1}[i],\cdots,\alpha_{r}[i]$,
and a corresponding non-zero value
of the $Y$, which is represented as the amplitude $f[i]$ in a list
\be
L^{(r)}[i]
=
\{
(\alpha_{1}[i],\cdots,\alpha_{r}[i]),
x_{\rm s}[i],x_{\rm e}[i];
f[i]
\}.
\ee
The number of lists $N_r$
depends on the expansion order $r$,
and of course on the shape of the parallel transporter
which one considers.
At this point the reader might think $x_{\rm s}$ and $x_{\rm e}$
are redundant but as we will see these are important
when dealing with the multiplication of lists.
A set of lists of order $r$,
$S^{(r)} = \{L^{(r)}[i]|i=1,...,N_r \}$,
is ``equivalent to'' the $r$'th order reduced vertex,
\be
\sum_{\alpha_1,...,\alpha_r}
Y_{\alpha_1,...,\alpha_r}
q_{\alpha_1}^{a_1} \cdots q_{\alpha_r}^{a_r}
=
\sum_{i=1}^{N_r}
q_{\alpha_{1}[i]}^{a_1} \cdots q_{\alpha_{r}[i]}^{a_r}
f[i],
\label{eqn:Yqqfqq}
\ee
for a fixed color configuration $\{a_1,\cdots,a_r\}$.
Then the parallel transporter in eq.(\ref{eqn:P}) is re-written as
\be
P({\cal L},x_{\rm s},x_{\rm e})
=
1+\sum_{r=1}^{\infty}
\frac{(ag_0)^r}{r!}
\sum_{a_1,\cdots,a_r}
{\cal C}^{a_1\cdots a_r}
\sum_{i=1}^{N_r}
q^{a_1}_{\alpha_{1}[i]} \cdots q^{a_r}_{\alpha_{r}[i]}
f[i].
\label{eqn:PP}
\ee
This expression is useful when deriving the multiplication algorithm
as we shall see soon.
A set\footnote{$S^{(r)}$ is a subset of $S$.}
$S=\{S^{(r)}|r=0,1,...,r_{\rm max}\}$,
which represents the reduced vertex up to the order $r_{\max}$
for the parallel transporter,
is now encoded in a program.

As an explicit example for the list, let us consider
the case of a single link variable.
From the expanded form
\be
U(x,\mu)
=
1+
\sum_{r=1}^{\infty}
\frac{(ag_0)^r}{r!}
\sum_{a_1,\cdots,a_r}
{\cal C}^{a_1 \cdots a_r}
q_{\alpha}^{a_1}\cdots q_{\alpha}^{a_r},
\label{eqn:UnonSF}
\ee
a corresponding list reads
\be
L^{(r)}
=
\{
(\underbrace{\alpha,\cdots,\alpha}_{\mbox{$r$ terms}}),
x,x+a\hat\mu;
1\},
\hspace{8mm}
\alpha=(\mu,x).
\label{eqn:linklist}
\ee
It consists of only one line ($N_r=1$)
and is the elementary building block of the algorithm,
that is, something like an initial condition.

So far we have defined the fundamental elements (the list and set)
on which the algorithm operates.
Next we discuss how to generate
the set $S$ for any parallel transporter.
First we assume that two sets
$S({\cal L}^{(n)},x_{\rm s}^{(n)},x_{\rm e}^{(n)})$ ($n=1,2$)
for parallel transporters
$P({\cal L}^{(n)},x_{\rm s}^{(n)},x_{\rm e}^{(n)})$
are known, and their elements are given by
\be
L^{(r)}[i]({\cal L}^{(n)},x_{\rm s}^{(n)},x_{\rm e}^{(n)})
=
\{
(\alpha_{1}^{(n)}[i],\cdots,\alpha_{r}^{(n)}[i]),
x_{\rm s}^{(n)},x_{\rm e}^{(n)};
f^{(n)}[i]
\}.
\ee
Then, we consider a product of the parallel transporters
\be
P({\cal L}^{(1)}*{\cal L}^{(2)},x_{\rm s}^{(1)},x_{\rm e}^{(2)}+\Lambda)
=
P({\cal L}^{(1)},x_{\rm s}^{(1)},x_{\rm e}^{(1)})
P({\cal L}^{(2)},x_{\rm s}^{(2)}+\Lambda,x_{\rm e}^{(2)}+\Lambda),
\label{eqn:MultP}
\ee
where $\Lambda=x_{\rm e}^{(1)}-x_{\rm s}^{(2)}$ is a shift vector
and is chosen such that the end point of the $n=1$ parallel transporter
and the starting point of the $n=2$ one are identical.
The shift for the $n=2$ parallel transporter in eq.(\ref{eqn:MultP}) is
taken into account for the list structure as
\bea
L^{(r)}[i]({\cal L}^{(2)},x_{\rm s}^{(2)},x_{\rm e}^{(2)})
&\stackrel{\mbox{\scriptsize $\Lambda$ shift}}{\longrightarrow}&
L^{(r)}[i]({\cal L}^{(2)},x_{\rm s}^{(2)}+\Lambda,x_{\rm e}^{(2)}+\Lambda)
\non\\
&=&
\{
(\tilde\alpha_{1}^{(2)}[i],\cdots,\tilde\alpha_{r}^{(2)}[i]),
x_{\rm s}^{(2)}+\Lambda,x_{\rm e}^{(2)}+\Lambda;
f^{(2)}[i]
\},
\eea
where all $x$ are shifted by $\Lambda$ and
$\tilde\alpha=(\mu,x+\Lambda)$.
The question now is how to obtain the set
for the product of the parallel transporter in eq.(\ref{eqn:MultP}),
\be
S({\cal L}^{(1)},x_{\rm s}^{(1)},x_{\rm e}^{(1)})
*
S({\cal L}^{(2)},x_{\rm s}^{(2)},x_{\rm e}^{(2)})
\longrightarrow
S({\cal L}^{(1)}*{\cal L}^{(2)},x_{\rm s}^{(1)},x_{\rm e}^{(2)}+\Lambda),
\ee
from the two given sets
$S({\cal L}^{(n)},x_{\rm s}^{(n)},x_{\rm e}^{(n)})$ with $n=1,2$.
Since any parallel transporter is composed of
the elementary single link variables,
by starting from a set for the link variable whose elements
are shown in eq.(\ref{eqn:linklist}) and by repeating the
above multiplication one can obtain a set for arbitrary parallel transporters.
This is the origin of the name  ``bottom up''.
The algorithm can be understood by looking at the actual
multiplication of two parallel transporters.
The right hand side of eq.(\ref{eqn:MultP}) is expanded as
\bea
\lefteqn{
P({\cal L}^{(1)},x_{\rm s}^{(1)},x_{\rm e}^{(1)})
P({\cal L}^{(2)},x_{\rm s}^{(2)}+\Lambda,x_{\rm e}^{(2)}+\Lambda)}\non\\
&=&
1+\sum_{r=1}^{\infty}
\frac{(ag_0)^r}{r!}
\sum_{a_1,\cdots,a_r}
{\cal C}^{a_1\cdots a_s}
{\cal C}^{a_{s+1}\cdots a_r}
\non\\
&\times&
\sum_{s=0}^{r}
\frac{r!}{s!(r-s)!}
\sum_{i=1}^{N_{s}^{(1)}}
\sum_{j=1}^{N_{r-s}^{(2)}}
q_{\alpha_{1}^{(1)}[i]}^{a_1}\cdots
q_{\alpha_{s}^{(1)}[i]}^{a_s}
q_{\tilde\alpha_{1}^{(2)}[j]}^{a_{s+1}}\cdots
q_{\tilde\alpha_{r-s}^{(2)}[j]}^{a_r}
f^{(1)}[i]f^{(2)}[j]
\non\\
&=&
1+\sum_{r=1}^{\infty}
\frac{(ag_0)^r}{r!}
\sum_{a_1,\cdots,a_r}
{\cal C}^{a_1\cdots a_r}
\sum_{k=1}^{N_r}
q_{\alpha_{1}[k]}^{a_1}\cdots
q_{\alpha_{s}[k]}^{a_s}
q_{\tilde\alpha_{s+1}[k]}^{a_{s+1}}\cdots
q_{\alpha_{r}[k]}^{a_r}
f[k].
\label{eqn:multP}
\eea
After the first equal-sign we inserted eq.(\ref{eqn:PP}).
In the last step,
we have used the fact that the color factor is independent of the shape of
the parallel transporter and
have combined the three summations
(over $s$, $i$ and $j$)
into that over $k$.
In short, we did a relabelling of the indices,
and rewrote the amplitude factor.
Finally the resulting lists
$L^{(r)}[k]({\cal L}^{(1)}*{\cal L}^{(2)},
x_{\rm s}^{(1)},x_{\rm e}^{(2)}+\Lambda)$
are created by putting
the new label structure and the new amplitude.
The algorithm is summarized as follows
\bi
\item Relabelling:
\be
\{\alpha_{1}^{(1)}[i],\cdots,\alpha_{s}^{(1)}[i],
\tilde\alpha_{1}^{(2)}[j],\cdots,\tilde\alpha_{r-s}^{(2)}[j]\}
\longrightarrow
\{\alpha_{1}[k],\cdots,\alpha_{s}[k],
\alpha_{s+1}[k],\cdots,\alpha_{r}[k]\},
\ee

\item Amplitude part:
\be
\frac{r!}{s!(r-s)!}f^{(1)}[i]f^{(2)}[j]\longrightarrow f[k],
\ee

\item Creating list:
\bea
\lefteqn{L^{(s)}[i]({\cal L}^{(1)},x_{\rm s}^{(1)},x_{\rm e}^{(1)})*
L^{(r-s)}[j]({\cal L}^{(2)},x_{\rm s}^{(2)}+\Lambda,x_{\rm e}^{(2)}+\Lambda)
\longrightarrow}
\non\\
&&
L^{(r)}[k]({\cal L}^{(1)}*{\cal L}^{(2)},
x_{\rm s}^{(1)},x_{\rm e}^{(2)}+\Lambda)
=
\{
(\alpha_{1}[k],\cdots,\alpha_{r}[k]),
x_{\rm s}^{(1)},x_{\rm e}^{(2)}+\Lambda;
f[k]\}.
\eea
\ei
This procedure should be carried out for $0 \le s \le r$,
$1 \le i \le N_s^{(1)}$ and $1 \le j \le N_{r-s}^{(2)}$
if order $r$ is desired.
This is an algorithm to obtain the set
$S({\cal L}^{(1)}*{\cal L}^{(2)},x_{\rm s}^{(1)},x_{\rm e}^{(2)}+\Lambda)$
and has been implemented in the python script language,
which has a great capability of dealing with complicated list operations.
In this way,
one can obtain the vertices
for any parallel transporter.

\section{Extension to the Schr\"odinger functional}
\label{sec:SF}
Let us now proceed to the SF~\cite{Luscher:1992}.
This section is the central part of the paper.

\subsection{Preliminary}
\label{sec:preliminary}
We consider, in the following section,
a finite box of size $L^3\times T$ with periodic boundary conditions in the
spatial directions.
We impose Dirichlet boundary
conditions for the link variables
at the time boundaries,
\begin{equation}
\left.U(x,k)\right|_{x_4 = 0} 
= 
\exp \{ a C \} ,\mbox{\hspace{5mm}}
\left.U(x,k)\right|_{x_4 = T} 
= 
\exp \{ a C^{\prime} \},
\label{eqn:CC}
\end{equation}
where the boundary fields $C$, $C^{\prime}$
in the SU(3) gauge theory~\cite{Luscher:1993gh}
are given in Appendix \ref{sec:BF}.
The SF in the theory is defined as
\begin{equation}
{\cal Z}[C,C^{\prime}]
=
\int D[U] e^{ - S[U] }.
\end{equation}

It is shown in Ref.~\cite{Luscher:1992} that a minimum of the
plaquette gauge action is given by the
lattice background field $U(x,\mu)=V(x,\mu)$, whose
elements are given by
\begin{equation}
V(x,4) = 1,
\mbox{\hspace{3mm}}   
V(x,k)
= 
V(x_4)=e^{ia({\cal E}x_4-iC)},
\label{eqn:V}
\end{equation}
where $V(x,k)$ is independent of the spatial direction $k$.
An explicit form of
the color electric field ${\cal E}$ is given in Appendix \ref{sec:BF}.
For later use, we introduce a convenient notation for
the background field
\bea
V(x,\mu)
&=&
(V(x_4))^{\sigma_{\mu}}
=
\left\{
\begin{array}{lll}
V(x_4)^{0}=1,      &\mbox{ for } \mu=4,    \\
V(x_4)^{1}=V(x_4), &\mbox{ for } \mu=1,2,3,
\end{array}
\right.
\\
\sigma_{\mu}
&=&
1-\delta_{\mu4}
=(1,1,1,0).
\label{eqn:sigma}
\eea
By using these notations,
we can express the classical link variable as
\be
V(x,\mu)^{m}
=
V(x_4)^{m\sigma_{\mu}},
\label{eqn:VVV}
\ee
where $m$ takes the values $\pm 1$ and $V(x,\mu)^{m}$ represents
$V(x,\mu)$ for $m=1$, and $V(x,\mu)^{\dag}=V(x,\mu)^{-1}$ for $m=-1$.

\subsection{An expression for the link variable}
\label{subsec:notations}
\begin{table}[t] 
\begin{center} 
\begin{tabular}{ccccc}
\hline \hline
$a$ & $\phi_a(x_4)$
    & $\phi_a^{\prime}$
    & $\pal_{\eta}\phi_a(x_4)$
    & $\pal_{\eta}\phi_a^{\prime}$
\\ \hline
1   & $-3 a\gamma x_4 + \frac{a}{L} ( \eta [\frac{3}{2} - \nu] - \frac{\pi}{3})$   
    & $-3 a^2\gamma$
    & $-3 ax_4/LT + \frac{a}{L}[\frac{3}{2} - \nu]$
    & $-3 a^2/LT$
\\
4   & $-3 a\gamma x_4 + \frac{a}{L} ( \eta [\frac{3}{2} + \nu] - \frac{2\pi}{3})$
    & $-3 a^2\gamma$
    & $-3 ax_4/LT + \frac{a}{L}[\frac{3}{2} + \nu]$
    & $-3 a^2/LT$
\\
3   & $0$
    & $0$
    & $0$
    & $0$
\\
6   & $\frac{a}{L}(2 \eta \nu - \frac{\pi}{3})$
    & $0$ 
    & $\frac{a}{L}2 \nu$ 
    & $0$ 
\\
8   & $0$
    & $0$
    & $0$
    & $0$
\\
\hline \hline
\end{tabular}
 \caption{
Phases $\phi_a(x_4)$, $\phi_a^{\prime}$, $\partial_{\eta}\phi_a(x_4)$
and $\partial_{\eta}\phi_a^{\prime}$ for SU(3) case.
The other components $a=2,5,7$, which are not shown here, are defined as
$\phi_2=-\phi_1$, $\phi_5=-\phi_4$, $\phi_7=-\phi_6$.
The $\gamma$ is given in eq.(\ref{eqn:gamma}) in Appendix \ref{sec:BF}.
}
\label{tab:phis}
\end{center}
\end{table}

In this section, we introduce a proper expression
for the link variable $U(x,\mu)$ in the SF case
which will be useful when generating vertices.
In the presence of a background field,
the link variable is written as
\be
U(x,\mu)
=
e^{a g_0 q_{\mu}(x)}V(x,\mu).
\label{eqn:link}
\ee
When dealing with the SF in the presence of
the abelian background field,
it is convenient to use a color decomposition,
\bea
q_{4}(x)
&=&
\sum_{a=1}^{8}q_{4}^{a}(x)I^a,
\label{eqn:q0}
\\
q_{k}(x)
&=&
\sum_{a=1}^{8}q_{k}^{a}(x)I^a
e^{i\phi_a(x_4)/2},
\label{eqn:qk}
\eea
where an overview of the phases $\phi_a(x_4)$ are given
in Table \ref{tab:phis}.
The idea of inclusion of the phase
can be found in the original SF paper~\cite{Luscher:1992}.
The basis of the Lie algebra $I^a$
can be found  in~\cite{Weisz:1996,Kurth:2004}
and is also summarized in
Appendix \ref{sec:liealgebra}.
By using $\sigma_{\mu}$ in eq.(\ref{eqn:sigma}),
the expression for
$q_{4}(x)$ and $q_{k}(x)$
are unified as
\bea
q_{\mu}(x)
&=&
\sum_{a=1}^{8}q_{\mu}^{a}(x)I^a
e^{i\sigma_{\mu}\phi_a(x_4)/2}.
\label{eqn:colordecomposition}
\eea
By using the above expression,
the link variable in eq.(\ref{eqn:link}) can be written as
\be
U(x,\mu)
=
V(x_4)^{\sigma_{\mu}}+
\sum_{r=1}^{\infty}
\frac{(ag_0)^r}{r!}
\sum_{a_1,\cdots,a_r}
q_{\alpha}^{a_1}
\cdots
q_{\alpha}^{a_r}
\left[
I^{a_1}\cdots I^{a_r}
V(x_4)^{\sigma_\mu}
\right]
e^{i\sigma_{\mu}\sum_{u=1}^r\phi_{a_{u}}(x_4)/2},
\label{eqn:linkphase}
\ee
where we have used eq.(\ref{eqn:VVV}) for the
lattice background field and
introduced the notation
$\left[ I I\cdots I \right]$
for representing a $3\times 3$ color matrix.

Now we will derive an expression for $U(x,\mu)^{\dag}=U(x,\mu)^{-1}$.
Thanks to the phase factor introduced in eq.(\ref{eqn:linkphase}),
it will turn out that it has a form similar to $U$.
Let us show this explicitly.
The inverse link variable can be expanded in the following way
\bea
U(x,\mu)^{-1}
&=&
V(x,\mu)^{-1}
e^{-ag_0q_{\mu}(x)}
=
V(x_4)^{-\sigma_{\mu}}
\non\\&+&
\sum_{r=1}^{\infty}
\frac{(-ag_0)^r}{r!}
\sum_{a_1,\cdots,a_r}
q_{\alpha}^{a_1}
\cdots
q_{\alpha}^{a_r}
\left[
V(x_4)^{-\sigma_\mu}
I^{a_1} \cdots I^{a_r}
\right]
e^{i\sigma_{\mu}\sum_{u=1}^r\phi_{a_{u}}(x_4)/2}.
\label{eqn:linkdag}
\eea
This naive form is not suited for our purpose.
The location of the $V$ is different from the one
in eq.(\ref{eqn:linkphase}).
We want to put $V$ to very right in the color matrix.
For this purpose,
we employ key formulas
which will play an important role in the following
\bea
\left[
V(x_4)I^aV(x_4)^{-1}
\right]
&=&
\left[
I^a
\right]
e^{i\phi_a(x_4)},
\label{eqn:VIV}
\\
\left[
e^{ia^2{\cal E}}
I^a
e^{-ia^2{\cal E}}
\right]
&=&
\left[I^a\right]
e^{i\phi_a^{\prime}},
\label{eqn:EIE}
\eea
where $\phi_a^{\prime}$ are shown in Table \ref{tab:phis}.
The formulas are derived from a special property of the
Abelian background field whose generators are
elements of the Cartan sub-algebra.
More details on the formulas are given
in Appendix \ref{sec:cartan}.
Let us return to $U^{-1}$.
By applying the formula in eq.(\ref{eqn:VIV})
to eq.({\ref{eqn:linkdag}}),
we obtain the following expression
\bea
U(x,\mu)^{-1}
&=&
V(x_4)^{-\sigma_{\mu}}
\non\\&+&
\sum_{r=1}^{\infty}
\frac{(-ag_0)^r}{r!}
\sum_{a_1,\cdots,a_r}
q_{\alpha}^{a_1}
\cdots
q_{\alpha}^{a_r}
\left[
I^{a_1}\cdots I^{a_r}
V(x_4)^{-\sigma_\mu}
\right]
e^{-i\sigma_{\mu}\sum_{u=1}^r\phi_{a_{u}}(x_4)/2},
\eea
where we have been able to place $V$
in very right in the color matrix.
Note the sign of the phase.
Now, we can express
the $U$ and $U^{-1}$ in a unified form,
for $m=\pm 1$,
\bea
U(x,\mu)^{m}
&=&
V(x_4)^{m\sigma_{\mu}}
\non\\&+&
\sum_{r=1}^{\infty}
\frac{(mag_0)^r}{r!}
\sum_{a_1,\cdots,a_r}
q_{\alpha}^{a_1}
\cdots
q_{\alpha}^{a_r}
\left[
I^{a_1}\cdots I^{a_r}
V(x_4)^{m\sigma_{\mu}}
\right]
e^{im\sigma_{\mu}\sum_{u=1}^r\phi_{a_{u}}(x_4)/2}.
\label{eqn:linkall}
\eea
Note the location of $m$.

In the following the term 'color factor' is reserved for
the color matrix part multiplied with the phase factors.
The corresponding color factor for the
single link variable reads
\be
\left[
I^{a_1}I^{a_2} \cdots I^{a_r}
V(x_4)^{m\sigma_{\mu}}
\right]
e^{im\sigma_{\mu}\sum_{u=1}^r\phi_{a_{u}}(x_4)/2}.
\label{eqn:color1}
\ee
An essential new ingredient
for the case of the SF is this color factor
which depends on $x_4$, $m$ and $\mu$, while
a color factor with the zero background field case
$\left[I^{a_1}I^{a_2} \cdots I^{a_r}\right]$
is independent of them.

\subsection{Multiplication of the color factors}
\label{subsec:multcolor}
Before entering the case of the SF,
it is maybe worth discussing the color factor
in the vanishing background field case
in order to make the difference between the two cases clear.
As mentioned in Section \ref{sec:bottomup},
the color factor in the vanishing background field case
is independent of the shape of the path,
therefore
the multiplication of color factors is trivial
\be
{\cal C}^{a_1\cdots a_{s}}
{\cal C}^{a_{s+1}\cdots a_{r}}
=
{\cal C}^{a_1\cdots a_s a_{s+1}\cdots a_{r}}.
\label{eqn:CCCmultzeroBGG}
\ee
On the other hand, for the SF case,
due to the presence of the non-vanishing background field,
the color factor depends on the shape of the path and
its multiplication is not as trivial as shown above.
In this section we will present
a multiplication rule for the color factor for the SF.

Our main finding is
that any color factor in the SF
of order $r$ can be cast in the form
\be
{\cal C}^{a_1\cdots a_r}(x_{l,4},A,B,C,D)
=
\underbrace{
\left[
I^{a_1} \cdots I^{a_r}
V(x_{l,4})^{A}
e^{ia^2{\cal E}B}
\right]}_{3\times 3 \mbox{ matrix}}
\underbrace{
e^{\frac{i}{2}\sum_{u=1}^r
(C_u\phi_{a_{u}}^{\prime}
+D_u\phi_{a_{u}}(x_{l,4}))}}_{U(1) \mbox{ phase}}.
\label{eqn:colorgeneral}
\ee
This can be shown by making use of two properties,
first, the background field being abelian
eq.(\ref{eqn:VIV},\ref{eqn:EIE}),
and second, the linear time dependence\footnote{
In some special case like~\cite{PerezRubio:2007qz},
the property of the linear time dependence is lost.
A multiplication algorithm for this case is given
in Appendix \ref{sec:timeV}, and it has already been
used in a one-loop calculation in the reference.
}
of the exponent of $V(x_4)$ in eq.(\ref{eqn:V}).
Note that the color factor has a time dependence
which comes from the background field.
We choose it as $x_{l,4}$ which is
the time component of the coordinate
of the left-most link variable of the parallel transporter
\be
P({\cal L},x_{\rm s},x_{\rm e})
=
U(x_{l},\mu)^m U U\cdots,
\ee
with $x_{\rm s}=x_{l}+a\hat{\mu}(1-m)/2$.
In practice, $x_{l}$ is set to the origin.
The components $A$ and $B$ are single component integer.
$C$ and $D$ are multi-component integers
of size $r$
\bea
C&=&(C_1,\cdots,C_r),
\\
D&=&(D_1,\cdots,D_r).
\eea
The $\phi_a^{\prime}$ in eq.(\ref{eqn:colorgeneral})
are the time derivative of the $\phi_a(x_4)$
and do them-self not depend on the time.
Their values are given in Table \ref{tab:phis}.
In the expression of the color factor,
the information about the lattice size
and the background field parameters ($\eta$, $\nu$) is encoded in
$V(x_{l,4})$, ${\cal E}$, and the phases
$\phi_a(x_{l,4})$, $\phi_a^{\prime}$.
When producing lists,
this information is actually irrelevant,
only $x_{l,4}$, $A$, $B$, $C$ and $D$ are required.
The former is only needed when writing down a numerical
expression of a vertex to a diagram calculation program
at a second stage.
The benefit of the expression is that
we can separate the information $x_{l,4}$, $A$, $B$, $C$ and $D$
(which will be included into the list structure)
and the lattice structure information.
Therefore, we can perform a symbolic list operation,
independently of the detail of the lattice size and
the background field parameters.

Let us give an example of a configuration of
$(x_{l,4},A,B,C,D)$ for the single link $U(x,\mu)^m$
whose color factor is given in eq.(\ref{eqn:color1})
\bea
x_{l,4}
&=&
x_4,
\\
A
&=&
m\sigma_{\mu},
\label{eqn:colorinitial1}
\\
B
&=&
0,
\\
C
&=&
(\underbrace{0,\cdots,0}_{\mbox{$r$ terms}}),
\\
D
&=&
(\underbrace{m\sigma_{\mu},\cdots,m\sigma_{\mu}}_{\mbox{$r$ terms}}).
\label{eqn:colorinitial2}
\eea

We have obtained a manageable expression for the color factor.
Next we formulate a multiplication for
the integer list $(x_{l,4},A,B,C,D)$.
An important point is that even in the SF
the multiplication of the color factor
is closed\footnote{This can be understood from the
background field gauge transformation in Ref.~\cite{Wolff:2007}}.
\bea
{\cal C}^{a_1\cdots a_s}(x_{l,4}^{(1)},A^{(1)},B^{(1)},C^{(1)},D^{(1)})
&\times&
{\cal C}^{a_{s+1}\cdots a_{r}}(x_{l,4}^{(2)},A^{(2)},B^{(2)},C^{(2)},D^{(2)})
\non\\
&=&
{\cal C}^{a_1\cdots a_r}(x_{l,4},A,B,C,D).
\eea
From an actual multiplication of the color factors,
we find the algorithm to get
a corresponding list from two given lists
$(x_{l,4}^{(1)},A^{(1)},B^{(1)},C^{(1)},D^{(1)})$ and
$(x_{l,4}^{(2)},A^{(2)},B^{(2)},C^{(2)},D^{(2)})$
in the following way
\bea
x_{l,4}
&\longleftarrow&
x_{l,4}^{(1)},
\label{eqn:multx4}
\\
A
&\longleftarrow&
A^{(1)}+A^{(2)},
\\
B
&\longleftarrow&
B^{(1)}+B^{(2)} + \Delta t A^{(2)},
\\
C
&\longleftarrow&
(
\underbrace{
\underbrace{C_{1}^{(1)},\cdots,C_{s}^{(1)}}_{\mbox{$s$ terms}},
\underbrace{C_{1}^{(2)}+2 B^{(1)}+\Delta
      t D_{1}^{(2)},
\cdots,
C_{r-s}^{(2)}+2 B^{(1)}+\Delta t D_{r-s}^{(2)}
}_{\mbox{$(r-s)$ terms}}
}_{\mbox{$r$ terms}}
),
\\
D
&\longleftarrow&
(
\underbrace{
\underbrace{D_{1}^{(1)},\cdots,D_{s}^{(1)}}_{\mbox{$s$ terms}},
\underbrace{D_{1}^{(2)}+2 A^{(1)},
\cdots,
D_{r-s}^{(2)}+2 A^{(1)}}_{\mbox{$(r-s)$ terms}}
}_{\mbox{ $r$ terms}}
),
\label{eqn:multD}
\eea
where $\Delta t=x_{l,4}^{(2)}-x_{l,4}^{(1)}$.
It turns out that the resulting $A$ and $B$ remain single component integers,
on the other hand, the resulting $C$ ($D$) is given by
combining $C^{(1)}$ ($D^{(1)}$) and $C^{(2)}$ ($D^{(2)}$)
with some additional terms for the latter part.
Since $x_{l,4}$, $A$, $B$, $C$ and $D$ are all integer values
and this operation is simple,
the multiplication algorithm is suited for a symbolic operation and is
easily implemented in the python script language.

In an actual implementation, 
as a new ingredient,
we have to add the new components $x_{l}$, $A$, $B$, $C$ and $D$
to the earlier list structure,
\bea
L^{(r)}[k]
&=&
\{
(\alpha_{1}[k],\cdots,\alpha_{r}[k]),
x_{\rm s},x_{\rm e},
;f[k]\}
\non\\
&\rightarrow&
\{
(\alpha_{1}[k],\cdots,\alpha_{r}[k]),
x_{\rm s},x_{\rm e},x_{l},
A[k],B[k],
\non\\
&&
(C_{1}[k],\cdots,C_{r}[k]),
(D_{1}[k],\cdots,D_{r}[k]);
f[k]
\}.
\eea
For two given lists, $n=1,2$,
\bea
L^{(r)}[i]({\cal L}^{(n)},x_{\rm s}^{(n)},x_{\rm e}^{(n)})
&=&
\{
(\alpha^{(n)}_{1}[i],\cdots,\alpha^{(n)}_{r}[i]),
x_{\rm s}^{(n)},x_{\rm e}^{(n)},x_{l}^{(n)},
A^{(n)}[i],B^{(n)}[i],
\non\\
&&
(C^{(n)}_{1}[i],\cdots,C^{(n)}_{r}[i]),
(D^{(n)}_{1}[i],\cdots,D^{(n)}_{r}[i]);
f^{(n)}[i]\},
\eea
the multiplication algorithm for lists is summarized as follows
(the relabelling and amplitude parts are the same as before).
\bi
\item Relabelling:
\be
\{\alpha_{1}^{(1)}[i],\cdots,\alpha_{s}^{(1)}[i],
\tilde\alpha_{1}^{(2)}[j],\cdots,\tilde\alpha_{r-s}^{(2)}[j]\}
\longrightarrow
\{\alpha_{1}[k],\cdots,\alpha_{s}[k],
\alpha_{s+1}[k],\cdots,\alpha_{r}[k]\},
\ee

\item Color factor part (eq.(\ref{eqn:multx4})-(\ref{eqn:multD})):
\bea
(x_{l,4}^{(1)},A^{(1)}[i],B^{(1)}[i],C^{(1)}[i],D^{(1)}[i])
&*&
(x_{l,4}^{(2)},A^{(2)}[j],B^{(2)}[j],C^{(2)}[j],D^{(2)}[j])
\non\\
&\longrightarrow&
(x_{l,4},A[k],B[k],C[k],D[k]),
\eea

\item Amplitude part:
\be
\frac{r!}{s!(r-s)!}f^{(1)}[i]f^{(2)}[j]\longrightarrow f[k],
\ee

\item Creating the list:
\bea
\lefteqn{
L^{(s)}[i]({\cal L}^{(1)},
x_{\rm s}^{(1)},x_{\rm e}^{(1)})*
L^{(r-s)}[j]({\cal L}^{(2)},
x_{\rm s}^{(2)}+\Lambda,x_{\rm e}^{(2)}+\Lambda)
}
\non\\
&\longrightarrow&
L^{(r)}[k]({\cal L}^{(1)}*{\cal L}^{(2)},
x_{\rm s}^{(1)},x_{\rm e}^{(2)}+\Lambda)
\non\\
&=&
\{
(\alpha_{1}[k],\cdots,\alpha_{r}[k]),
x_{\rm s}^{(1)},x_{\rm e}^{(2)}+\Lambda,
x_{l},A[k],B[k],
\non\\
&&
(C_{1}[k],\cdots,C_{r}[k]),
(D_{1}[k],\cdots,D_{r}[k]);
f[k]\}.
\eea
\ei

In order to start the multiplication algorithm,
we need an initial set of lists.
It is the set for a single link variable $U(x,\mu)^m$
which is given by the list
\be
L^{(r)}
=
\{
(\underbrace{\alpha,\cdots,\alpha}_{\mbox{$r$ terms}}),
x_{\rm s},x_{\rm e},x_{l},
m\sigma_{\mu},0,
(\underbrace{0,\cdots,0}_{\mbox{$r$ terms}}),
(\underbrace{m\sigma_{\mu},\cdots,m\sigma_{\mu}}_{\mbox{$r$ terms}});
m^r
\},
\ee
with
\bea
x_{\rm s}
&=&
x+a\hat{\mu}\frac{1-m}{2},
\\
x_{\rm e}
&=&
x+a\hat{\mu}\frac{1+m}{2},
\\
x_{l}
&=&
x.
\eea

To check that the algorithm works properly,
we perform a one-loop calculation of the SF coupling.
The results are shown in Section \ref{sec:application}.
By calculating this quantity,
we can check the two-point vertex.
We investigate not only the plaquette gauge action
but also the improved gauge actions including six-link loops.
As a further check we compute the Big Mac diagram, which includes
three-point vertex,
for smaller lattices and check
the consistency with the old results
(private communication with Peter Weisz).

\section{How to exploit the set for the gauge action}
\label{sec:WritingVertex}
In this section,
by using the python output (set $S$) generated by the above algorithm,
we describe how to implement the vertex
in a code which calculates Feynman diagrams.
We exclusively consider the gauge actions, and
in both coordinate space and time-momentum space
for future purposes.
In this section we write $\mu$ and $x$ explicitly
instead of using the super index $\alpha=(\mu,x)$,
and set the lattice unit $a=1$.

\subsection{Coordinate space representation}
\label{sec:vertexp}
The gauge action\footnote{We will give more precise definition of the
gauge action in Section \ref{sec:application}.
Here the reality and trace properties are important.}
consists of a loop ${\cal C}$
\be
S_{\rm G}
=
\frac{2}{g_0^2}\sum_{{\cal C}}
{\rm Re} {\rm tr} [1 - U({\cal C})],
\ee
where $U({\cal C})$ is the parallel transporter around ${\cal C}$.
Its expansion in terms of $g_0$ in the coordinate space
is given by
\bea
S_{\rm G}
&=&
\sum_{r=2}^{\infty}\frac{(g_0)^{r-2}}{r!}
\sum_{x_1,\cdots,x_r}
\sum_{\mu_1,\cdots,\mu_r}
\sum_{a_1,\cdots,a_r}
V^{a_1 \cdots a_r}_{\mu_1 \cdots \mu_r}
(x_1,\cdots,x_r)
\prod_{j=1}^r q^{a_j}_{\mu_j}(x_j).
\eea
The symmetrized vertex $V$ is given by summation over
unsymmetrized vertices $Y$
\be
V^{a_1 \cdots a_r}_{\mu_1 \cdots \mu_r}
(x_1,\cdots,x_r)
=
\frac{1}{r!}
\sum_{\sigma\in{\cal S}_r}
\sigma \cdot Y^{a_1 \cdots a_r}_{\mu_1 \cdots \mu_r}
(x_1,\cdots,x_r),
\ee
where ${\cal S}_r$ is the permutation group of the order $r$.
The symmetrized vertex may be rewritten as
\be
V^{a_1 \cdots a_r}_{\mu_1 \cdots \mu_r}
(x_1,\cdots,x_r)
=
\frac{1}{r!}
\sum_{\sigma\in{\cal S}_r/{\cal Z}_r}
\sigma \cdot Y^{\prime a_1 \cdots a_r}_{\mu_1 \cdots \mu_r}
(x_1,\cdots,x_r),
\label{eqn:VSF}
\ee
where the partially symmetrized vertex is
\be
Y^{\prime a_1 \cdots a_r}_{\mu_1 \cdots \mu_r}
(x_1,\cdots,x_r)
=
\sum_{\sigma\in{\cal Z}_r}
\sigma \cdot Y^{a_1 \cdots a_r}_{\mu_1 \cdots \mu_r}
(x_1,\cdots,x_r).
\label{eqn:Yprime}
\ee
${\cal Z}_r$ is defined as a subset
of the total permutation group ${\cal S}_r$ and consists of
the inversion and cyclic permutations.
The summation over ${\cal Z}_r$ in eq.(\ref{eqn:Yprime})
can be done at the python level even for the SF.
Some details about the partial symmetrization is
given in Appendix \ref{sec:symmetrization}.
The number of elements of ${\cal S}_r$ and ${\cal Z}_r$
are $\#{\cal S}_r=r!$ and $\#{\cal Z}_r=2r$ respectively.
Therefore there are remaining permutations
$\#{\cal S}_r/{\cal Z}_r=r!/2r=(r-1)!/2$.
Note that for $r\ge4$ we need to perform
the remaining symmetrization (summation in eq.(\ref{eqn:VSF})) at the
Feynman diagram calculation stage.

To make the discussion more concrete,
we use the plaquette gauge action as a specific example.
First we consider a sum of the plaquette over all directions
at fixed point
\be
\sum_{\mu>\nu}{\rm Retr}\left[
U(0,\mu)U(0+\hat\mu,\nu)U(0+\hat\nu,\mu)^{\dag}U(0,\nu)^{\dag}
\right],
\label{eqn:Pzero}
\ee
and the corresponding set whose elements are lists of
the from
\bea
L^{(r)}[i]
&=&
\{ (\mu_{1}[i],\cdots,\mu_{r}[i]),
   ( dx_{1}[i],\cdots, dx_{r}[i]),
0,0,0,
A[i],B[i],
\non\\
&&
(C_{1}[i],\cdots,C_{r}[i]),(D_{1}[i],\cdots,D_{r}[i]);f[i]
\},
\eea
where we assume that the partial symmetrization procedure has been done.
As described in Ref.~\cite{Luscher:1985wf},
by making use of the translation invariance in the
spatial directions,
the $d{\bf x}_1,...,d{\bf x}_r$
are translated to the standard form~\cite{Hart:2004bd},
so that the number of lists can be reduced.
On the other hand, this reduction does not apply for the time direction
due to the loss of the time translation invariance in the SF.
Note that in eq.(\ref{eqn:Pzero})
the summation over the coordinates is not taken but kept.
In this case we may set $x_{\rm s}=x_{\rm e}=x_{l}=0$.
If one carries out the sum at the python level,
it renders the number of lists prohibitively large of order $L^3T$,
and this is
apparently inefficient.
Instead the sum is carried out in the diagram calculation code.
In order to obtain the complete vertex,
we have to loop over the $x_{l}$ exhausting the whole elements
of the vertex.
In this way, all non-zero elements of the vertex are assigned.

\begin{figure}[t]
 \begin{center}
   \scalebox{0.6}{\includegraphics{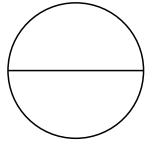}}
 \end{center}
 \caption{The Big Mac diagram
\label{fig:diagrams}}
\end{figure}

Let us show how to use the the set for  the partially
symmetrized vertex in a diagram calculation code.
As an example, let us take
the Big Mac diagram, which appears in the two-loop
SF coupling calculation \cite{Narayanan:1995ex},
shown in Figure \ref{fig:diagrams},
whose expression in the coordinate space is given by
\bea
{\rm BigMac}
&=&
\frac{1}{12}
\sum_{\{a\},\{\mu\},\{x\}}
V^{a_1a_2a_3}_{\mu_1\mu_2\mu_3}(x_1,x_2,x_3)
V^{a_4a_5a_6}_{\mu_4\mu_5\mu_6}(x_4,x_5,x_6)
\non\\
&&\times
\langle q^{a_1}_{\mu_1}(x_1) q^{a_4}_{\mu_4}(x_4) \rangle_0
\langle q^{a_2}_{\mu_2}(x_2) q^{a_5}_{\mu_5}(x_5) \rangle_0
\langle q^{a_3}_{\mu_3}(x_3) q^{a_6}_{\mu_6}(x_6) \rangle_0,
\label{eqn:BM}
\eea
where $1/12$ is a symmetry factor and $V$ is the symmetrized
three-point gluon vertex.
The free gluon propagator in
the coordinate space is expressed in terms of that
in the time-momentum space $D^a_{\mu\nu}(\pbf,x_0,y_0)$,
which gauge fixing is done,
\be
\langle q^{a}_{\mu}(x) q^{b}_{\nu}(y) \rangle_0
=
\delta_{b\bar{a}}
S^a_{\mu\nu}(x_0,y_0,\xbf,\ybf),
\ee
and
\bea
S^a_{\mu\nu}(x_0,y_0,\xbf,\ybf)
&=&
S^a_{\mu\nu}(x_0,y_0,\xbf-\ybf)
\non\\
&=&
\frac{1}{L^3}
\sum_{\pbf}
e^{i\pbf(\xbf-\ybf+\frac{1}{2}\hat\mu\sigma_\mu
                  -\frac{1}{2}\hat\nu\sigma_\nu)}
D^a_{\mu\nu}(\pbf,x_0,y_0).
\label{eqn:propFourier}
\eea
The notation $\bar{a}$ is defined in Appendix \ref{sec:liealgebra}.
In the following, we use the notation $t=x_4$.
By using the propagator $S$ and performing the partial sum over
the color index $a_4$, $a_5$ and $a_6$,
the BigMac diagram is given by
\bea
{\rm BigMac}
&=&
\frac{1}{12}
\sum_{a_1,a_2,a_3}
\sum_{\{\mu\},\{t\},\{\xbf\}}
V^{a_1a_2a_3}_{\mu_1\mu_2\mu_3}
(t_1,t_2,t_3,\xbf_1,\xbf_2,\xbf_3)
V^{\bar{a_1}\bar{a_2}\bar{a}_3}_{\mu_4\mu_5\mu_6}
(t_4,t_5,t_6,\xbf_4,\xbf_5,\xbf_6)
\non\\
&&\times
S^{a_1}_{\mu_1\mu_4}(t_1,t_4,\xbf_1-\xbf_4)
S^{a_2}_{\mu_2\mu_5}(t_2,t_5,\xbf_2-\xbf_5)
S^{a_3}_{\mu_3\mu_6}(t_3,t_6,\xbf_3-\xbf_6).
\label{eqn:BigMac}
\eea
The color factor of order $r$ for the gauge action is given by
\bea
\lefteqn{{\cal C}_{{\rm G}}^{a_1\cdots a_r}(t,A,B,C,D)}
\non\\
&=&
{\rm tr}[
I^{a_1}\cdots I^{a_r}e^{ i{\cal E}B}
+(-1)^r
I^{a_r}\cdots I^{a_1}e^{-i{\cal E}B}
]
e^{\frac{i}{2}\sum_{u=1}^r(C_{u} \phi_{a_u}^{\prime}+D_{u}
\phi_{a_u}(t))}.
\eea
Actually almost all elements of the color factor are zero,
therefore it is convenient to use a color index
\be
(a_1[n],a_2[n],a_3[n]),
\ee
with running index $n$ which contains indices for
the nonzero elements of the color factor.
Furthermore, another list structure is used for the vertex.
The symmetrized vertex of order $r=3$ is written as
\be
V^{a_1a_2a_3}_{\mu_1\mu_2\mu_3}(x_1,x_2,x_3)
=
\frac{1}{3!}
Y^{\prime a_1a_2a_3}_{\mu_1\mu_2\mu_3}(x_1,x_2,x_3),
\ee
where $Y^{\prime}$ is the partially
symmetrized vertex.
At order $r=3$, $V$ and $Y^{\prime}$ are equivalent
up to an overall factor $3!$.
The partially symmetrized vertex can be generated from the python
output, the list structure for the first gluon vertex
in eq.(\ref{eqn:BigMac})
is generated from
\bea
L^{(3)}[i]=
\{
(\mu_1[i],\mu_2[i],\mu_3[i]),
((dt_1[i],d\xbf_1[i]),(dt_2[i],d\xbf_2[i]),(dt_3[i],d\xbf_3[i])),
\non\\
0,0,0,A[i],B[i],
(C_1[i],C_2[i],C_3[i]),
(D_1[i],D_2[i],D_3[i]);
f[i]
\},
\label{eqn:pythonlist}
\eea
with index $i$ identifying the list.
The number of lists is
$N_3=396$ for the plaquette action.
For the second vertex
we use the index $j$ for identification.
The coordinates $t_m$ and $\xbf_m$ ($m=1,2,3,4,5,6$) are given
in terms of the elements of the list as well as $t$, $s$, $\xbf$ and $\ybf$,
\begin{alignat}{2}
\xbf_1&=\xbf+d\xbf_1[i]&,
\hspace{10mm}
t_1&=t+dt_1[i],
\label{eqn:x1t1i}
\\
\xbf_2&=\xbf+d\xbf_2[i]&,
\hspace{10mm}
t_2&=t+dt_2[i],
\\
\xbf_3&=\xbf+d\xbf_3[i]&,
\hspace{10mm}
t_3&=t+dt_3[i],
\\
\xbf_4&=\ybf+d\xbf_1[j]&,
\hspace{10mm}
t_4&=s+dt_1[j],
\\
\xbf_5&=\ybf+d\xbf_2[j]&,
\hspace{10mm}
t_5&=s+dt_2[j],
\\
\xbf_6&=\ybf+d\xbf_3[j]&,
\hspace{10mm}
t_6&=s+dt_3[j].
\label{eqn:x6t6j}
\end{alignat}
Now the indices ($t$, $\xbf$) and ($s$, $\ybf$) run over the whole lattice.
By making use of the above parameterizations and the
translation invariance for the spatial direction,
the diagram is written as
\bea
{\rm BigMac}
&=&
\frac{L^3}{12}
\sum_{n(\mbox{\tiny color loop})}^{52}
\sum_{i,j(\mbox{\tiny python loop})}^{396}
\sum_{t,s}
\sum_{\xbf}
\non\\
&&\phantom{\times}
\frac{1}{3!}
f[i]{\cal C}_{\rm G}^{a_1[n]a_2[n]a_3[n]}
(t,A[i],B[i],C[i],D[i])
\non\\
&&\times
\frac{1}{3!}
f[j]{\cal C}_{\rm G}^{\bar{a}_1[n]\bar{a}_2[n]\bar{a}_3[n]}
(s,A[j],B[j],C[j],D[j])
\non\\
&&\times
S^{a_1[n]}_{\mu_1[i]\mu_1[j]}(t+dt_1[i],s+dt_1[j],\xbf+d\xbf_1[i]-d\xbf_1[j])
\non\\
&&\times
S^{a_2[n]}_{\mu_2[i]\mu_2[j]}(t+dt_2[i],s+dt_2[j],\xbf+d\xbf_2[i]-d\xbf_2[j])
\non\\
&&\times
S^{a_3[n]}_{\mu_3[i]\mu_3[j]}(t+dt_3[i],s+dt_3[j],\xbf+d\xbf_3[i]-d\xbf_3[j]).
\eea
Note that the summations over $\{\mu\}$, $\{t\}$ and $\{\xbf\}$
in eq.(\ref{eqn:BM})
of order $4^6\times T^6(L^3)^6$
are replaced by $i,j$ (python loop), $t,s$ and $\xbf$
of order $(396)^2\times T^2L^3$.
In the summation step, one has to be cautious about the boundary conditions,
\be
q_k(x)|_{x_4=0}=
q_k(x)|_{x_4=T}=0.
\ee

Here we give a short comment
about the color factor of the gauge action.
For any gauge action composed of
closed and traced loops,
it holds that $A=0$.
This means that the color matrix part
(which is the color factor without the phase factor
that depends on $x_4$),
\be
{\rm tr}
\left[
I^{a_1}\cdots I^{a_r}
e^{i{\cal E}B}
+(-1)^r
I^{a_r}\cdots I^{a_1}
e^{-i{\cal E}B}
\right],
\label{eqn:colormatrixG}
\ee
is time independent.
Therefore one does not have to
calculate the color matrix for every $x_4$
when writing down the vertex
in computer code at the Feynman diagram calculation stage.
Before the time loop $x_4$,
one can store the color matrix in somewhere,
and then in the time loop it can be called.
This is realized by virtue of separating the
color factor into two parts,
the information about lists and the background field.

\subsection{Time-momentum space representation}
The Fourier transformation for the 
quantum field is given by
\be
q^{a}_{\mu}(x)
=
\frac{1}{L^3}
\sum_{{\bf p}}e^{i{\bf p}\cdot ({\bf x}+\frac{\hat{\mu}}{2}\sigma_{\mu})}
\tilde{q}^{a}_{\mu}({\bf p},x_4),
\ee
where the spatial momenta are given by
$p_k=2\pi n_k/L$ with $n_k=0,...,L-1$.
An expansion of the action in terms of the coupling $g_0$ in the
time-momentum space is given by
\bea
S_{\rm G}
&=&
\sum_{r=2}^{\infty}\frac{(g_0)^{r-2}}{r!}
\left( \frac{1}{L^3}\right)^{r-1}
\sum_{{\bf p}_1,\cdots,{\bf p}_r}
\delta_{{\bf p}_1+\cdots+{\bf p}_r,2\pi{\bf n}}
\non\\
&&
\sum_{\mu_1,\cdots,\mu_r}
\sum_{t_1,\cdots,t_r}
\sum_{a_1,\cdots,a_r}
V^{a_1 \cdots a_r}_{\mu_1 \cdots \mu_r}
({\bf p}_1,\cdots,{\bf p}_r;t_1,\cdots,t_r)
\prod_{j=1}^r \tilde{q}^{a_j}_{\mu_j}(-{\bf p}_j,t_j).
\label{eqn:SGTM}
\eea

Similar way to the case in the coordinate space,
the symmetrized vertex is given by
the partially symmetrized vertex
\be
V^{a_1 \cdots a_r}_{\mu_1 \cdots \mu_r}
({\bf p}_1,\cdots,{\bf p}_r;t_1,\cdots,t_r)
=
\frac{1}{r!}
\sum_{\sigma\in{\cal S}_r/{\cal Z}_r}
\sigma \cdot Y^{\prime a_1 \cdots a_r}_{\mu_1 \cdots \mu_r}
({\bf p}_1,\cdots,{\bf p}_r;t_1,\cdots,t_r).
\label{eqn:permutationsum}
\ee
We also have written python code for the time-momentum representation.
The list structure for the time-momentum space
is a little bit different from the one in the coordinate space.
The differences occur in the coordinate of the link variables only,
$x\rightarrow (t,{\bf v})$, where $t=x_{4}$ and ${\bf v}$
is a spatial component of mid-point of the link variables in two lattice units.
The list for eq.(\ref{eqn:Pzero}) is given by
\bea
L^{(r)}[i]&=&
\{ (\mu_{1}[i],\cdots,\mu_{r}[i]),
   ( (t_{1}[i],{\bf v}_{1}[i]),\cdots,(t_{r}[i],{\bf v}_{r}[i]),
0,0,0,
\non\\
&&
A[i],B[i],(C_{1}[i],\cdots,C_{r}[i]),(D_{1}[i],\cdots,D_{r}[i]);f[i]
\}.
\eea

By making use of the propagator in the time-momentum space
\be
\langle
\tilde{q}^{a}_{\mu}({\bf p},x_0)
\tilde{q}^{b}_{\nu}({\bf q},y_0)
\rangle_0
=
\delta_{b\bar{a}}L^3\delta_{\pbf+\qbf,\zerobf}
D^a_{\mu\nu}(\pbf,x_0,y_0),
\ee
the Big Mac diagram in the space is written as
\bea
{\rm BigMac}
&=&
\frac{1}{12L^3}
\sum_{n(\mbox{\tiny color loop})}^{52}
\sum_{i,j(\mbox{\tiny python loop})}^{396}
\sum_{t,s}
\sum_{\pbf_1,\pbf_2}
\non\\
&&\phantom{\times}
\frac{1}{3!}
f[i]{\cal C}_{\rm G}^{a_1[n]a_2[n]a_3[n]}
(t,A[i],B[i],C[i],D[i])
e^{-\frac{i}{2}
({\bf p}_1\cdot{\bf v}_{1}[i]
+{\bf p}_2\cdot{\bf v}_{2}[i]
-{\bf q}\cdot{\bf v}_{3}[i])}
\non\\
&&\times
\frac{1}{3!}
f[j]{\cal C}_{\rm G}^{\bar{a}_1[n]\bar{a}_2[n]\bar{a}_3[n]}
(s,A[j],B[j],C[j],D[j])
e^{ \frac{i}{2}
({\bf p}_1\cdot{\bf v}_{1}[j]
+{\bf p}_2\cdot{\bf v}_{2}[j]
-{\bf q}\cdot{\bf v}_{3}[j])}
\non\\
&&\times
D^{\bar{a}_1[n]}_{\mu_1[j]\mu_1[i]}(\pbf_1,t+dt_1[j],s+dt_1[i])
\non\\
&&\times
D^{\bar{a}_2[n]}_{\mu_2[j]\mu_2[i]}(\pbf_2,t+dt_2[j],s+dt_2[i])
\non\\
&&\times
D^{      a_3[n]}_{\mu_3[i]\mu_3[j]}(\qbf,t+dt_3[i],s+dt_3[j]),
\eea
where $q_k=2\pi m_k/L$ with $m_k=n_{1k}+n_{2k} \mbox{ mod } L$
for $p_{ik}=2\pi n_{ik}/L$ ($i=1,2$).
We have used the relation,
\be
D^a_{\mu\nu}(\pbf,x_0,y_0)
=
D^{\bar{a}}_{\nu\mu}(-\pbf,y_0,x_0).
\ee
Note that the computational effort
scales with $T^2L^6$ which is more demanding
compared to the coordinate space by $L^3$.
Even though the cost for
the Fourier transformation
of the propagator in eq.(\ref{eqn:propFourier})
scales proportional to $T^2L^6$,
the prefactor is negligible compared to the
sum in the diagram whose factor is $52\times (396)^2$.
This shows the benefit of using the coordinate space
for the computation of bubble-type diagrams \cite{Narayanan:1995ex}.

\section{Applications}
\label{sec:application}
In this section, we show some applications of the
algorithm, especially we consider the one-loop
computation of the SF coupling.
From this computation we can check that
the algorithm works properly
and furthermore we can investigate
more complicated gauge actions,
like improved gauge actions including
chair type and parallelogram loops.

\subsection{Improved gauge actions}
\begin{figure}[t]
\begin{center}
      \psfragscanon
      \psfrag{c0}[][][2]{$c_0$}
      \psfrag{c1}[][][2]{$c_1$}
      \psfrag{c2}[][][2]{$c_2$}
      \psfrag{c3}[][][2]{$c_3$}
      \scalebox{0.62}{\includegraphics{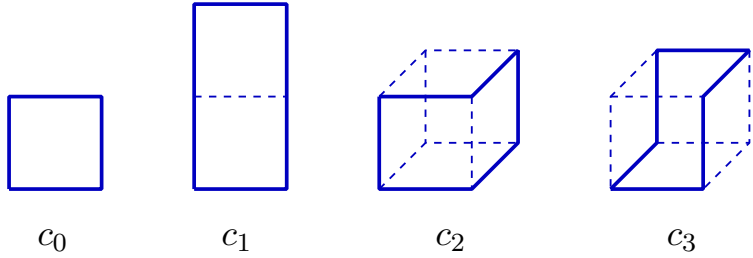}}
\caption{Loops together with corresponding factors.
\label{fig:loop}}
\end{center}
\end{figure}

In this subsection, we summarize the definition
of the improved gauge actions including six-link loops.
More details about the definition can be found in~\cite{Aoki:1998qd}.
The improved gauge action for the SF is given by
\be
S_{\rm G}
=
\frac{2}{g_0^2}\sum_{i=0}^{3}\sum_{{\cal C}\in {\cal S}_i}
W_i({\cal C},g_0^2)
{\rm Retr}\left[1-U({\cal C})\right],
\ee
where $U({\cal C})$ is a parallel transporter along a loop ${\cal C}$.
Since we are taking the real part,
loops ${\cal C}$ which differ by orientation only are considered equal.
${\cal S}_i$ denotes sets of loops ${\cal C}$ on the lattice
as given in Figure \ref{fig:loop} together with
the corresponding factors $c_i$, which
are normalized by $c_0+8c_1+16c_2+8c_3=1$.
We adopt the weight factors for ``Choice B'' in Ref.~\cite{Aoki:1998qd}.
For the plaquette, this means
\be
  W_0({\cal C},g^2_0) 
  = 
  \left\{
   \begin{array}{ll}
    c_{\rm{s}}(g^2_0)
    &\mbox{if ${\cal C}$ lie on one of the boundaries,}\\
    c_0 c^P_{\rm{t}}(g^2_0)
    &\mbox{if ${\cal C}$ just touch one of the boundaries,}\\
    c_0
    &\mbox{otherwise,}
   \end{array}
  \right.
  \label{eqn:W0}
\ee
and for rectangular loops
\be
  W_1({\cal C},g^2_0)
  =
  \left\{
  \begin{array}{ll}
   0
   & \mbox{if ${\cal C}$ lie completely on one of the boundaries,}\\
   c_1 c^R_{\rm{t}}(g^2_0)
   &\mbox{if ${\cal C}$ have exactly 2 links on a boundaries, }\\
   c_1
   &\mbox{otherwise,}
  \end{array}
  \right.
  \label{eqn:W1}
\ee
with the perturbative expansion of the
boundary counter term~\cite{Takeda:2003he}
\bea
 c_0 c^P_{\rm{t}}(g^2_0)
  & = & 
  c_0 ( 1 + c^{P(1)}_{\rm{t}} g^2_0 + O(g^4_0) ), \\
 c_1 c^R_{\rm{t}}(g^2_0)
  & = & 
  c_1 ( 3/2 + c^{R(1)}_{\rm{t}} g^2_0 + O(g^4_0) ).
\eea
The tree values of the boundary counter terms
were given in Ref.~\cite{Aoki:1998qd}.
The one-loop coefficients for the improved
gauge actions composed from plaquette loops and
rectangular loops
(Iwasaki, Symanzik and DBW2\footnote{Note that the value of $c_1$ for
the DBW2 here is a bit
different from that in~\cite{Takeda:2003he}.})
were already determined in Ref.~\cite{Takeda:2003he} with a constraint
\be
c_{\rm t}^{R(1)}
=
2c_{\rm t}^{P(1)}.
\label{eqn:cr2cp}
\ee
As discussed in the reference,
this constraint is convenient when
we deal with one-loop O($a$) improvement,
and we assume this in the following.
As we shall see later,
the remaining one-loop values $c_{\rm t}^{P(1)}$
will be fixed
for the other improved gauge actions including
the chair and parallelogram loops.
The weight factors
for the chair and parallelogram loops ($i=2,3$)
are given by
\be
 W_i({\cal C},g^2_0) 
  = 
  \left\{
   \begin{array}{ll}
    0          
    & \mbox{if ${\cal C}$ lie completely on one of the boundaries,}\\
    c_i
    & \mbox{otherwise. }
   \end{array}
  \right.
\ee

Gauge actions including the chair and parallelogram loops
are hardly used in current simulations,
but here we consider such actions for comparison
with the old perturbation theory results.
We deal with seven gauge actions whose
name and values of $c_i$ are given in Table \ref{tab:A0A1}.

\subsection{One-loop results for the SF coupling}
We perform the one-loop computation
of the SF coupling in pure SU(3) gauge theory
for the improved gauge actions.
We set $T=L$ as usual.
The SF coupling is defined~\cite{Luscher:1993gh} by
\be
\bar{g}^2_{\rm SF}(L)
=
k
\left.\left(
\frac{\pal \Gamma}{\pal \eta}
\right)^{-1}
\right|_{\eta=\nu=0},
\ee
with $\Gamma=-\ln {\cal Z}[C,C^{\prime}]$ and a normalization
\be
k
=
12(L/a)^2
\left[
             c_0\{\sin(2\gamma)+\sin( \gamma)\}
+4(c_1+2c_2+c_3)\{\sin(4\gamma)+\sin(2\gamma)\}
\right],
\ee
where $\gamma$ is given in eq.({\ref{eqn:gamma}}).

The coupling constant has the perturbative expansion
\be
\bar{g}^2_{\rm SF}(L)
=
g_0^2+m_1^{(1)}(L/a)g_0^4+O(g_0^6),
\label{eqn:gSFg0}
\ee
and the one-loop coefficient is given by
\bea
m_1^{(1)}(L/a)
&=&
m_1^{(0)}(L/a)
-2 c_{\rm t}^{P(1)} R(L/a) a/L,
\label{eqn:m11}
\\
R(L/a)
&=&
\frac
{c_0\{\sin(2\gamma)+\sin(\gamma)\}+4 c_1          \{\sin(4\gamma)+\sin(2\gamma)\}}
{c_0\{\sin(2\gamma)+\sin(\gamma)\}+4(c_1+2c_2+c_3)\{\sin(4\gamma)+\sin(2\gamma)\}}.
\eea
$m_1^{(0)}(L/a)$ is the one-loop coefficient
which is computed with the tree level value of the boundary
counter terms, that is, the
tree level O($a$) improved coefficient.
On the other hand, $m_1^{(1)}(L/a)$ is a one-loop O($a$) improved one
which is expected to have no $a/L$ term.
What we compute actually is $m_1^{(0)}(L/a)$
in a range $L/a=6,...,64$.
The contribution is separated into two parts,
the gauge field and the ghost parts,
\be
m_1^{(0)}=-\frac{1}{2k} \sum_{a,{\bf p}}
{\rm Tr}\left[
(K^{a}({\bf p}))^{-1}\frac{\pal K^{a}({\bf p})}{\pal \eta}
\right]
+ (\mbox{ghost contribution}).
\ee
The inverse propagator $K^{a}({\bf p})$ is
a sum of the two-point vertex $V^{ab}_{\mu\nu}({\bf p},-{\bf p};t,s)$
in eq.(\ref{eqn:SGTM})
and the conventional gauge fixing term~\cite{Luscher:1993gh}.
We use the ghost action in the previous reference.
The $\eta$ derivative of $K^{a}({\bf p})$
can be build numerically from the
python output as shown in Appendix \ref{sec:etaderivative}.
In the one-loop computation we adopt the time-momentum representation
which is still advantageous at this order.
The trace is taken over the Lorentz and time indices.
$c_{\rm P}^{(1)}$ in eq.(\ref{eqn:m11}) can be chosen to cancel
the linear term of $a/L$ in $m_1^{(0)}(L/a)$,
and then $m_1^{(1)}(L/a)$ turns out to be the O($a$)
improved quantity.

By analyzing the tree level O($a$) improved one-loop results
following~\cite{Bode:1999sm}, we obtain the
first several coefficients of the Symanzik expansion form,
\be
m_1^{(0)}(L/a)=A_0+B_0 \ln (a/L) +A_1 a/L + B_1 a/L \ln (a/L)+ O((a/L)^2).
\label{eqn:m1symanzik}
\ee
$B_0$ should be $2b_0$, which is the universal one-loop coefficient
of the beta function.
We check that $|B_0/(2b_0)-1|<10^{-4}$ holds for all improved gauge actions,
which we calculate.
We also observe that  $|B_1|<10^{-2}$ for all cases,
and $B_1=0$ is a signal for the achievement of the tree level
O($a$) improvement.
By assuming that $B_0=2b_0$ and $B_1=0$ we obtain the values
of $A_0$ and $A_1$, which are given in Table \ref{tab:A0A1}.
Their values for the case of the plaquette action~\cite{Luscher:1993gh}
(the explicit value of $A_0$ is shown in~\cite{Bode:1999sm}),
and rectangular type actions~\cite{Takeda:2003he},
(Iwasaki and Symanzik) are already known
and we confirmed the values to ensure the consistency.

\begin{table}[t]
\begin{center}
\begin{tabular}{|l|l|l|l|l|l|}
     \hline\hline
Action & \multicolumn{1}{c|} {$c_1$}
& \multicolumn{1}{c|} {$c_2$}
& \multicolumn{1}{c|}{$c_3$}
& \multicolumn{1}{c|}{$A_0$}
& \multicolumn{1}{c|}{$A_1$} \\
     \hline
Plaquette&$\pho 0$  &$\pho 0$   &$\pho 0$  
 &$\pho 0.3682819( 6)$&$-0.1779( 3)$    \\
   \hline
Wilson RG&$-0.252$  &$\pho 0$   &$-0.170$  
 &$-0.2184103( 8)$    &$\pho 0.2987( 3)$\\
   \hline
Iwasaki&$-0.331$  &$\pho 0$   &$\pho 0$  
 &$-0.204903( 3)$     &$\pho 0.304( 2)$ \\
   \hline
DBW2&$-1.4088$ &$\pho 0$   &$\pho 0$  
 &$-0.62816( 5)$      &$\pho 0.90( 2)$  \\
   \hline
Symanzik&$-1/12$   &$\pho 0$   &$\pho 0$  
 &$\pho 0.1361508( 3)$&$-0.0060( 1)$    \\
   \hline
Symanzik II&$-1/12$   &$\pho 1/16$&$-1/16$   
 &$\pho 0.1403119( 5)$&$\pho 0.0216( 2)$\\
   \hline
Symanzik III&$-1/12$   &$-0.1$     &$\pho 0.1$
 &$\pho 0.1491183( 6)$&$-0.0745( 3)$    \\
   \hline\hline
\end{tabular}
\end{center}
\caption{The coefficients $A_0$ and $A_1$ of asymptotic expansion
of the one-loop coefficient for the various gauge actions.
}
\label{tab:A0A1}
\end{table}

Let us discuss the ratio of the Lambda parameters.
Actually there are several old perturbation theory results
of the ratio of the lambda parameters between
the plaquette gauge action and the improved gauge actions,
therefore we can compare with our results.
Before taking a continuum limit, we have to perform the renormalization.
We introduce a renormalized coupling $\bar{g}_{\rm Lat}(\mu)$ through
\bea
g_0^2
&=&
\bar{g}_{\rm Lat}^2(\mu)
+z_1(a\mu)\bar{g}_{\rm Lat}^4(\mu)
+O(\bar{g}_{\rm Lat}^6(\mu)),
\label{eqn:g0glat}
\\
z_1(a\mu)
&=&
2b_0\ln(a\mu).
\eea
By substituting eq.(\ref{eqn:g0glat}) into
the expression of $\bar{g}_{\rm SF}^2(L)$ of eq.(\ref{eqn:gSFg0})
with eq.(\ref{eqn:m1symanzik}), we obtain
\be
\bar{g}_{\rm SF}^2(L)
=
\bar{g}_{\rm Lat}^2(\mu)
+
\bar{g}_{\rm Lat}^4(\mu)
\left[
A_0 +B_0 \ln(a/L)
+2b_0\ln(a\mu)
+O(a/L)
\right]
+O(\bar{g}_{\rm Lat}^6(\mu)).
\ee
By noting $B_0=2b_0$ and
setting $\mu=1/L$, we can cancel the log divergence.
Then we can take the continuum limit
\be
\bar{g}_{\rm SF}^2(L)
=
\bar{g}_{\rm Lat}^2(1/L)
+
A_0
\bar{g}_{\rm Lat}^4(1/L)
+O(\bar{g}_{\rm Lat}^6(1/L)).
\ee
The ratio of the lambda parameters is given in terms of $A_0$
\be
\frac{\Lambda_{\rm Lat}}{\Lambda_{\rm SF}}
=
\exp[-A_0/(2b_0)].
\ee
From this, we can compose a ratio
\be
\frac{\Lambda_{\rm impr}}{\Lambda_{\rm Plaq}}
=
\frac{\Lambda_{\rm impr}/\Lambda_{\rm SF}}{\Lambda_{\rm Plaq}/\Lambda_{\rm SF}}
=
\exp[(A_0^{\rm Plaq}-A_0^{\rm impr})/(2b_0)].
\ee
The resulting values are given in Table \ref{tab:ratio1},
we observe rough consistency with the old values.
Another ratio of the lambda parameters
in the ``lattice scheme'' and the $\overline{\rm MS}$ scheme is given by
\bea
\frac{\Lambda_{\rm Lat}}{\Lambda_{\overline{\rm MS}}}
&=&
\frac{\Lambda_{\rm Lat}}{\Lambda_{\rm SF}}
\frac{\Lambda_{\rm SF}}{\Lambda_{\overline{\rm MS}}}
=
\exp[-(A_0+c_1/(4\pi))/(2b_0)],
\\
c_1|_{N_{\rm f}=0}
&=&
1.255621(2),
\eea
where $c_1$ is taken from the two-loop SF coupling paper~\cite{Bode:1999sm},
its value for the several gauge actions is given in Table \ref{tab:ratio2}.
This table may be useful for future references.

\begin{table}[t]
 \begin{center} 
  \begin{tabular}{|c|lllllll|}
  \hline\hline
  Action(impr.)
& our results
& ~\cite{Skouroupathis:2007mq}
& ~\cite{Weisz:1983bn}
& ~\cite{Iwasaki:1983zm}
& ~\cite{Iwasaki:1984cj}
& ~\cite{Ukawa:1983ae}
& ~\cite{Bernreuther:1984wx}
\\ \hline
Wilson RG   &$67.4384(5)$&$-$      &$-$       &$67.97(9)$&$-$          &$67.6(3)$&$67.37(14)$\\ \hline
Iwasaki     &$61.207(2)$ &$61.2064$&$-$       &$-$       &$59.05(1.00)$&$-$      &$-$        \\ \hline
DBW2        &$1277.1(5)$ &$1276.44$&$-$       &$-$       &$-$          &$-$      &$-$        \\ \hline
Symanzik    &$5.29209(3)$&$5.29210$&$5.294(4)$&$5.29(1)$ &$-$          &$5.29(1)$&$5.29(1)$  \\ \hline
Symanzik II &$5.13636(3)$&$-$      &$-$       &$-$       &$-$          &$5.13(1)$&$-$        \\ \hline
Symanzik III&$4.82173(3)$&$-$      &$-$       &$4.842(2)$&$-$          &$-$      &$-$        \\
  \hline\hline
  \end{tabular}
\caption{Status of $\Lambda_{\rm impr}/\Lambda_{\rm Plaq}$
for pure SU(3) gauge theory.
The results from Ref.~\cite{Iwasaki:1983zm,Iwasaki:1984cj}
tend to disagree with ours.
Other values from~\cite{Skouroupathis:2007mq,Weisz:1983bn,Ukawa:1983ae,Bernreuther:1984wx} are consistent with our results.
Note that the value of $c_1$ in~\cite{Skouroupathis:2007mq} is
a little bit different from ours and actually they use $c_1=-1.4086$.
\label{tab:ratio1}}
 \end{center}
\end{table}

\begin{table}[t]
 \begin{center} 
  \begin{tabular}{|c|l|}
  \hline\hline
  Action
& $\Lambda_{\rm Lat}/\Lambda_{\overline{\rm MS}}$
\\ \hline
Plaquette   &$\pho 0.0347109675049892 (2)$ \\ \hline
Wilson RG   &$\pho 2.34086 (2 )$ \\ \hline
Iwasaki     &$\pho 2.12455(5)$   \\ \hline
DBW2        &$44.33 (2)$         \\ \hline
Symanzik    &$\pho 0.1836938(4)$ \\ \hline
Symanzik II &$\pho 0.1782883(4)$ \\ \hline
Symanzik III&$\pho 0.1673674(8)$ \\
  \hline\hline
  \end{tabular}
\caption{$\Lambda_{\rm Lat}/\Lambda_{\overline{\rm MS}}$ of SU(3) gauge theory.
For the plaquette gauge action it is taken from Ref.~\cite{Luscher:1995nr},
where a direct calculation of the ratio was done.
Our results are those for the other gauge actions.
\label{tab:ratio2}}
 \end{center}
\end{table}

From the resulting value of $A_1$ we can determine the one-loop coefficient
of the O($a$) boundary counter term $c_{\rm t}^{P(1)}$.
The improvement condition yields
\be
c_{\rm t}^{P(1)}=\frac{A_1}{2(c_0+8c_1)}.
\ee
Together with eq.(\ref{eqn:cr2cp}), one can achieve one-loop O($a$) improvement.

\subsection{Relative deviation of the step scaling function}
\begin{figure}[t]
 \begin{center}
  \begin{tabular}{cc}
  \scalebox{1.1}{\includegraphics{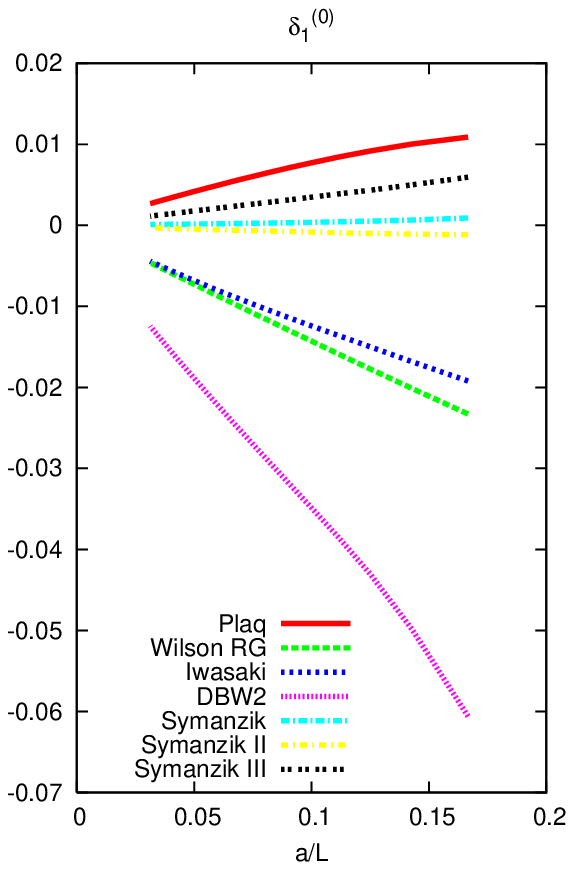}}&
  \scalebox{1.1}{\includegraphics{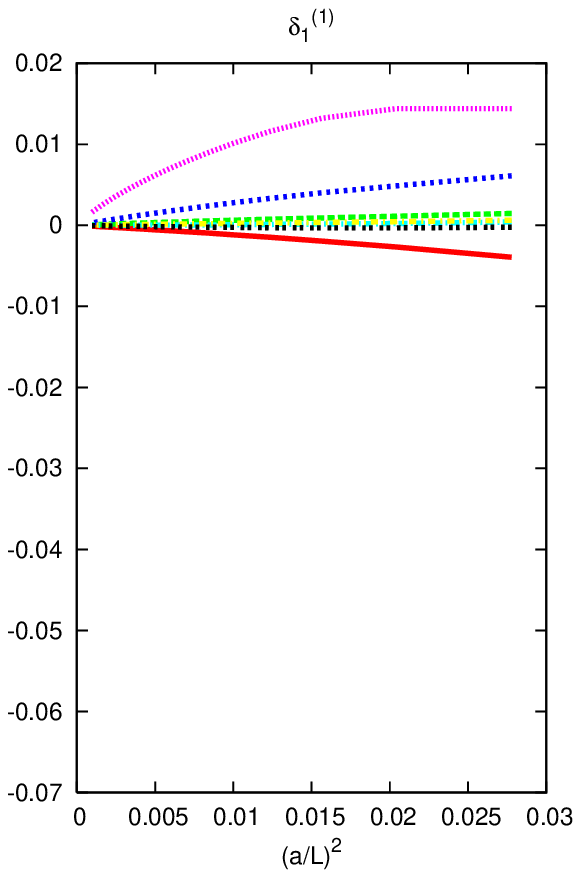}}
  \end{tabular}
 \end{center}
 \caption{Left panel is for the tree level O($a$) improved case,
and right one is for the one-loop relative deviation
where the O($a$) term is removed.
\label{fig:deviation}}
\end{figure}

Finally, we investigate the relative deviation of the step scaling
function (SSF) for the various gauge actions.
The SSF for the running coupling~\cite{Luscher:1991wu} is defined by
\be
\sigma(u)=\bar{g}_{\rm SF}^2(2L),
\hspace{5mm}
u=\bar{g}_{\rm SF}^2(L).
\ee
We denote the SSF measured on the lattice with $\Sigma(u,a/L)$.
It is expected to converge to the continuum value,
\be
\sigma(u)=\lim_{a\rightarrow 0}\Sigma(u,a/L).
\ee
A scaling behavior to the limit can be described by the
relative deviation which is defined from
\be
\delta(u,a/L)=\frac{\Sigma(u,a/L)-\sigma(u)}{\sigma(u)}.
\ee
In perturbation theory, it is expanded as
\be
\delta(u,a/L)=\delta_1(a/L)u +O(u^2).
\ee
The one-loop coefficient is given by
\be
\delta_1^{(k)}(a/L)=m_1^{(k)}(2L)-m_1^{(k)}(L/a)-2b_0 \ln 2,
\ee
where $k=0,1$ denotes the degree of improvement,
the results
are shown in Figure \ref{fig:deviation} for the various gauge actions.
Symanzik type actions have relatively good scaling behavior, while
the renormalization group improved  types show rather
large cutoff effects at one-loop level.

\section{Concluding remarks}
\label{sec:conclusion}
In this paper we derive the multiplication algorithm
to generate Feynman rules for
the SF with an abelian background field.
The essential new ingredient
for the extension is how to treat
the color factor which involves the background field.
By making use of the
key formulas eq.(\ref{eqn:VIV}) and eq.(\ref{eqn:EIE}),
which are derived from special properties
of the abelian background field (eq.(\ref{eqn:key})),
this difficulty is resolved.
We present the multiplication
algorithm for the color factors in eq.(\ref{eqn:multx4}-\ref{eqn:multD}).
By making use of the algorithm one can obtain
vertices to any order for any shape of the parallel transporter
in the SF.
One has to keep in mind that
as shown in Appendix \ref{sec:symmetrization}
one needs to perform
the remaining symmetrization for the vertices $r\le 4$ of the gauge action
in order to obtain the totally symmetrized vertex
at the Feynman diagram calculation stage.

As a first application and to check the
correctness of the python code,
we calculated the one-loop coefficient
of the SF coupling for various gauge actions.
We observe consistency by
comparing with previous calculations,
and analyze the scaling behavior of the step scaling functions.
As a further check we compute the Big Mac diagram
for smaller lattices and observe a consistency
with the old results.
As a further application, we applied the automatic method
to a lattice with $L=T \pm a$ which
is motivated by considering staggered fermions in the SF
~\cite{PerezRubio:2007qz}.
Actually, on that lattice, the time dependence of the background field
is not uniform anymore.
Therefore we extend our algorithm to this case,
the details of the extended algorithm are shown in Appendix \ref{sec:timeV}.
A preliminary result for the gauge sector by making use of the algorithm
was already presented in~\cite{PerezRubio:2007qz,Takeda:2007dt}.

It is desirable to go beyond one-loop for actual applications.
In that line, the full two-loop computation of the SF coupling is
the next target.
In this paper, we exclusively consider
the gluonic sector as a specific example.
Of course, however the multiplication algorithm
can be applied for the fermion actions,
for example, Wilson fermions and
also for the clover term.
The algorithm is suited for all link connected objects.
Therefore an application for the HQET with
highly smearing is also possible.

Some readers, who are interested to use the generalized
version of the python script, are invited to ask the author to send it.

\section*{Acknowledgments}
I would like to thank Alistair Hart
to provide me the original python script
which was the starting point of my study.
I am also grateful to Dirk Hesse, Georg von Hippel,
Rainer Sommer and Ulli Wolff
for discussions, critical reading and giving comments on the manuscript.
I appreciate Peter Weisz sending me the
numerical results of the Big Mac diagram.
This work is supported in the framework of SFB Transregio 9
of the Deutsche Forschungsgemeinschaft (DFG).
I also thank FLAVIAnet for financial support.

\begin{appendix}
\section{The basis of the Lie algebra $su(3)$}
\label{sec:liealgebra}
For the Lie algebra $su(3)$ one may choose as a basis
\be
T^a = \frac{1}{2i} \lambda^a,
\ee
with the Gell-Mann matrices $\lambda^a$
for $a=1,2,...,8$, such that
\be
{\rm tr}\left[T^a T^b \right]=-\frac{1}{2}\delta^{ab},
\hspace{5mm}
\left[T^a,T^b\right]=f^{abc}T^c.
\ee

For the SF case, however, it is convenient to choose another basis.
New matrices $\tilde{\lambda}^a$~\cite{Weisz:1996,Kurth:2004}
are introduced, which coincide with
the Gell-Mann matrices $\lambda^a$ except for the two
diagonal matrices,
\be
\tilde{\lambda}^3 = -\frac{1}{2}\lambda^3 +\frac{\sqrt{3}}{2}\lambda^8,
\ee
\be
\tilde{\lambda}^8 = \frac{\sqrt{3}}{2}\lambda^3 +\frac{1}{2}\lambda^8.
\ee
The matrices $\tilde{\lambda}^a$ are 
\bea
\tilde{\lambda}^1 = \left(
\begin{array}{ccc}
0 & 1 & 0 \\
1 & 0 & 0 \\
0 & 0 & 0 \\
\end{array}
\right), &\qquad &
\tilde{\lambda}^2 = \left(
\begin{array}{ccc}
0 & -i & 0 \\
i & 0 & 0 \\
0 & 0 & 0 \\
\end{array}
\right), \nonumber\\
\tilde{\lambda}^4 = \left(
\begin{array}{ccc}
0 & 0 & 1 \\
0 & 0 & 0 \\
1 & 0 & 0 \\
\end{array}
\right), &\qquad & 
\tilde{\lambda}^5 = \left(
\begin{array}{ccc}
0 & 0 & -i \\
0 & 0 & 0 \\
i & 0 & 0 \\
\end{array}
\right), \nonumber\\
\tilde{\lambda}^6 = \left(
\begin{array}{ccc}
0 & 0 & 0 \\
0 & 0 & 1 \\
0 & 1 & 0 \\
\end{array}
\right), &\qquad & 
\tilde{\lambda}^7 = \left(
\begin{array}{ccc}
0 & 0 & 0 \\
0 & 0 & -i \\
0 & i & 0 \\
\end{array}
\right), \nonumber\\
\tilde{\lambda}^3 = \left(
\begin{array}{ccc}
0 & 0 & 0 \\
0 & 1 & 0 \\
0 & 0 & -1 \\
\end{array}
\right), &\qquad& 
\tilde{\lambda}^8 = \frac{1}{\sqrt{3}}\left(
\begin{array}{ccc}
2 & 0 & 0 \\
0 & -1 & 0 \\
0 & 0 & -1 \\
\end{array}
\right).
\eea
After a normalization,
\be
\tilde T^a =\frac{1}{2i}\lambda^a,
\ee
these matrices may be used to define a new basis $I^a$, which is given by
\bea
I^1 = T_+ = \frac{1}{\sqrt{2}}(\tilde T^1+i\tilde T^2), &\qquad &
I^2 = T_- = \frac{1}{\sqrt{2}}(\tilde T^1-i\tilde T^2), \nonumber\\
I^4 = U_+ = \frac{1}{\sqrt{2}}(\tilde T^4+i\tilde T^5), &\qquad &
I^5 = U_- = \frac{1}{\sqrt{2}}(\tilde T^4-i\tilde T^5), \nonumber\\
I^6 = V_+ = \frac{1}{\sqrt{2}}(\tilde T^6+i\tilde T^7), &\qquad &
I^7 = V_- = \frac{1}{\sqrt{2}}(\tilde T^6-i\tilde T^7), 
\eea
for the non--diagonal matrices and
\be
I^3 = \tilde T^3, \qquad  I^8 =\tilde T^8,
\ee
for the diagonal ones. 
For this basis, one has
\be
I^{a\dagger} = -I^{\bar{a}},
\ee
where $\bar{1}=2$, $\bar{4}=5$, $\bar{6}=7$, and $\bar{2}=1$ and so on.
For the diagonal
matrices, one has $\bar{3}=3$ and $\bar{8}=8$. The normalization is chosen
such that
\be
{\rm tr} \left[I^a I^b\right] = -\frac{1}{2}\delta^{b\bar{a}}.
\ee

\section{The boundary fields and the constant color electric background field}
\label{sec:BF}

In the basis chosen in Appendix \ref{sec:liealgebra},
the boundary fields are expressed as
\bea
C
&=&
\frac{i}{L}
\left(
\begin{array}{ccc}
\eta-\frac{\pi}{3} &  0 & 0 \\
0 & \eta(-\frac{1}{2}+\nu) & 0 \\
0 &  0 & -\eta(\frac{1}{2}+\nu)+\frac{\pi}{3}\\
\end{array}
\right)
\non\\
&=&
\frac{i}{L}
\left[
 (\eta\nu-\frac{\pi}{6})\tilde{\lambda}^3
+\frac{\sqrt{3}}{2}(\eta-\frac{\pi}{3})\tilde{\lambda}^8
\right],
\\
C^{\prime}
&=&
\frac{i}{L}
\left(
\begin{array}{ccc}
-\eta-\pi &  0 & 0 \\
0 & \eta(\frac{1}{2}+\nu)+\frac{\pi}{3} & 0 \\
0 &  0 & \eta(\frac{1}{2}-\nu)+\frac{2\pi}{3}\\
\end{array}
\right)
\non\\
&=&
\frac{i}{L}
\left[
 (\eta\nu-\frac{\pi}{6})\tilde{\lambda}^3
-\frac{\sqrt{3}}{2}(\eta+\pi)\tilde{\lambda}^8
\right].
\eea

The constant color electric background field
$\mathcal{E}$ is proportional to $I^8$
\be
\mathcal{E} = -\gamma\left(
\begin{array}{ccc}
2 &  0 & 0 \\
0 & -1 & 0 \\
0 &  0 & -1\\
\end{array}
\right) = -\gamma\sqrt{3}\tilde{\lambda}^8
=
-2\sqrt{3}i\gamma I^8,
\ee
with
\be
\gamma = \frac{1}{LT}\left(\eta + \frac{\pi}{3}\right).
\label{eqn:gamma}
\ee
The $\eta$ derivative of ${\cal E}$ is given by
\be
\frac{\pal{\cal E}}{\pal\eta}
=
-2\sqrt{3}i\frac{\pal\gamma}{\pal\eta}I^8
=
\frac{-2\sqrt{3}i}{LT}I^8.
\label{eqn:etaE}
\ee

\section{The Cartan sub-algebra}
\label{sec:cartan}

Let us assume that $H_i$ for $i=1,\cdots,m$ ($m$ is the rank of the algebra)
are hermitian and elements of the Cartan sub-algebra.
They commute,
\be
[H_i,H_j]=0.
\ee
The generators $E_{a}$ of the original algebra satisfy
\be
[H_i,E_{a}]
=
\mu_{a i}
E_{a},
\label{eqn:ladder}
\ee
where $\mu_{a i}$ are the roots, which
are the weights of the adjoint representation.
By making use of eq.(\ref{eqn:ladder}),
one can see that
\be
e^{i\sum_{j}h_jH_j}
E_{a}
e^{-i\sum_{j}h_jH_j}
=
E_{a}
e^{i\sum_{j}h_j\mu_{a j}},
\label{eqn:key}
\ee
with real coefficients $h_i$.
An $E_a$ sandwiched between group elements generated by the Cartan algebra
does not mix with other $E_b$ ($b\ne a$),
but turns out to be the
$E_{a}$ itself multiplied a phase factor.

In our case of $su(3)$,
the Cartan generators are identified as $H_1=iI^3$, $H_2=iI^8$, and
the other generators as $E_a=iI^a$.
The values of the roots
for the $su(3)$ case are shown in Table \ref{tab:roots}.
In terms of the Cartan generator,
the background field in eq.(\ref{eqn:V}) can be written as
\be
V(x_4)=e^{i \sum_j h_j(x_4) H_j},
\ee
where the coefficients $h_j(x_4)$ can be read off from
the definition of $V(x_4)$.
By making use of the above formula,
one can see that the phase $\phi_a(x_4)$
in eq.(\ref{eqn:VIV}) is given
in terms of the coefficients $h_j$ and the roots $\mu_{aj}$
\be
\phi_a(x_4)=\sum_j h_j(x_4) \mu_{aj}.
\ee

\begin{table}[t] 
\begin{center} 
\begin{tabular}{ccccccccc}
\hline \hline
$a$ & 1 & 2 & 3 & 4 & 5 & 6 & 7 & 8 \\
\hline
$\mu_{a1}$ &$-\frac{1}{2}$ & $\frac{1}{2}$ & $0$ & $\frac{1}{2}$ &$-\frac{1}{2}$ & $1$ &$-1$ &$0$ \\
$\mu_{a2}$ &$ \frac{\sqrt{3}}{2}$ &$-\frac{\sqrt{3}}{2}$ & $0$ & $\frac{\sqrt{3}}{2}$ &$-\frac{\sqrt{3}}{2}$ & $0$ & $0$ &$0$ \\
\hline \hline
\end{tabular}
 \caption{The roots $\mu_{ai}$ for SU(3) case.}
\label{tab:roots}
\end{center}
\end{table}

\section{Multiplication for a background field with an arbitrary time dependence}
\label{sec:timeV}
\begin{figure}[!t]
\begin{center}
\psfragscanon
\psfrag{t}    [][][1.6]{$t$}
\psfrag{min}  [][][1.6]{$t_{\rm min}$}
\psfrag{max}  [][][1.6]{$t_{\rm max}$}
\psfrag{start}[][][1.6]{$t_{\rm start}$}
\psfrag{end}  [][][1.6]{$t_{\rm end}$}
\scalebox{0.6}{\includegraphics{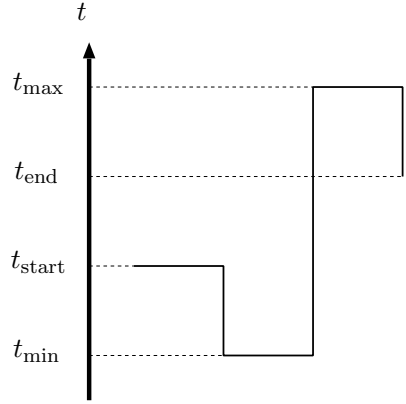}}
\end{center}
\caption{A line which show $t_{\rm max}$, $t_{\rm min}$, $t_{\rm start}$
and $t_{\rm end}$.
\label{fig:line}}
\end{figure}

In this section, we provide a multiplication algorithm for
the color factor for an abelian background field
which has arbitrary time dependence.
We start from a general form of the color factor of order $r$
for this case,
\bea
\lefteqn{{\cal C}^{a_1\cdots a_r}(t_{\rm min},t_{\rm max},A,D)}
\non\\
&=&
\left[
I_{a_1} \cdots I_{a_r}
V(t_{\rm min})^{A_{t_{\rm min}}}
V(t_{\rm min}+1)^{A_{t_{\rm min}+1}}
\cdots
V(t_{\rm max})^{A_{t_{\rm max}}}
\right]\times
\non\\
&&
\times
e^{\frac{i}{2}\sum_{u=1}^r
(D_{u,t_{\rm min}}\phi_{a_{u}}(t_{\rm min})
+D_{u,t_{\rm min}+1}\phi_{a_{u}}(t_{\rm min}+1)
+\cdots
+D_{u,t_{\rm max}}\phi_{a_{u}}(t_{\rm max})
)}.
\eea
For a certain configuration $\{a_1,\cdots,a_r\}$,
the color factor is specified by the
set $\{t_{\rm min},t_{\rm max},A,D\}$.
$t_{\rm min}$ and $t_{\rm max}$ are defined
in Figure \ref{fig:line}.
The explicit forms of $A$ and $D$ are given by
\bea
A
&=&
(A_{t_{\rm min}},A_{t_{\rm min}+1},\cdots,A_{t_{\rm max}}),
\\
D
&=&
(D_{t_{\rm min}},D_{t_{\rm min}+1},\cdots,D_{t_{\rm max}}),
\\
D_{t}&=&(D_{1t},D_{2t},\cdots,D_{rt}),
\mbox{ for }
t_{\rm min}\le t \le t_{\rm max}.
\eea
Note that all elements of $A$, $D$ are integer valued.
Let us see an example of
$(A,D)$ for the color factor for the single link variable $U(x,\mu)^m$.
For $x=0$ and $\mu=4$ ($t_{\rm min}=0$, $t_{\rm max}=1$),
\bea
(A,D)
&=&
((A_0,A_1),(D_0,D_1))
\non\\
&=&
((0,0),(\underbrace{(0,0,...,0)}_{\mbox{$r$ terms}},
\underbrace{(0,0,...,0)}_{\mbox{$r$ terms}})).
\label{eqn:single0}
\eea
For $x=0$ and $\mu=1,2,3$ ($t_{\rm min}=0$, $t_{\rm max}=0$),
\bea
(A,D)
&=&
(A_0,D_0)
\non\\
&=&
(0,\underbrace{(0,0,...,0)}_{\mbox{$r$ terms}}).
\label{eqn:singlek}
\eea

Even for this case, the color factors are closed under the
multiplication ($r=r_1+r_2$),
\bea
&&
{\cal C}^{a_1\cdots a_{r_1}}
(t_{\rm min}^{(1)},t_{\rm max}^{(1)},A^{(1)},D^{(1)})
{\cal C}^{a_{r_1+1}\cdots a_{r_1+r_2}}
(t_{\rm min}^{(2)},t_{\rm max}^{(2)},A^{(2)},D^{(2)})
\non\\ &=&
{\cal C}^{a_1\cdots a_r}(t_{\rm min},t_{\rm max},A,D).
\eea
So we found the following multiplication algorithm.
First, the resulting $t_{\rm min}$ and $t_{\rm max}$ are given by,
\bea
t_{\rm min}
&=&
\min (t_{\rm min}^{(1)},t_{\rm min}^{(2)}),
\label{eqn:tmin}
\\
t_{\rm max}
&=&
\max (t_{\rm max}^{(1)},t_{\rm max}^{(2)}).
\label{eqn:tmax}
\eea
Before operating $A$ and $D$, we have to pad zeros, from the left
\bea
A^{(1)}
\longrightarrow
\tilde A^{(1)}
&=&
(\tilde A_{t_{\rm min}}^{(1)},\cdots,\tilde A_{t_{\rm max}}^{(1)})
\non\\
&=&
\left\{
\begin{array}{cc}
(\underbrace{0,0,...,0}_{\mbox{$(t_{\rm min}^{(1)}-t_{\rm min})$ terms}},
A_{t_{\rm min}^{(1)}}^{(1)},...,A_{t_{\rm max}^{(1)}}^{(1)}),
&
\mbox{ for }
t_{\rm min}^{(1)} > t_{\rm min},
\\
(A_{t_{\rm min}^{(1)}}^{(1)},...,A_{t_{\rm max}^{(1)}}^{(1)}),
&
\mbox{ for }
t_{\rm min}^{(1)} = t_{\rm min},
\end{array}
\right.
\label{eqn:paddingmin}
\eea
where we note that the size of $\tilde A^{(1)}$ is $t_{\max}-t_{\min}+1$,
while that of $A^{(1)}$ is $t_{\max}^{(1)}-t_{\min}^{(1)}+1$.
Also from the right,
\bea
A^{(1)}
\longrightarrow
\tilde A^{(1)}
&=&
(\tilde A_{t_{\rm min}}^{(1)},\cdots,\tilde A_{t_{\rm max}}^{(1)})
\non\\
&=&
\left\{
\begin{array}{cc}
(A_{t_{\rm min}^{(1)}}^{(1)},...,A_{t_{\rm max}^{(1)}}^{(1)},
\underbrace{0,0,...,0}_{\mbox{$(t_{\rm max}-t_{\rm max}^{(1)})$ terms}}),
&
\mbox{ for }
t_{\rm max}^{(1)} < t_{\rm max},
\\
(A_{t_{\rm min}^{(1)}}^{(1)},...,A_{t_{\rm max}^{(1)}}^{(1)}),
&
\mbox{ for }
t_{\rm max}^{(1)} = t_{\rm max}.
\end{array}
\right.
\label{eqn:paddingmax}
\eea
In a similar way,  we can pad the others with zeros
and obtain $\tilde A^{(2)}$ and $\tilde D^{(1,2)}$.
Then the operation is done for the tilde objects,
\bea
A_t
&=&
\tilde A_{t}^{(1)}+\tilde A_{t}^{(2)},
\label{eqn:A}
\\
D_t
&=&
(
\underbrace{
\underbrace{\tilde D_{1t}^{(1)},\cdots,\tilde D_{r_1t}^{(1)}}_
{\mbox{$r_1$ terms}},
\underbrace{\tilde D_{1t}^{(2)}+2\tilde A_{t}^{(1)},
\cdots,
\tilde D_{r_2t}^{(2)}+2\tilde A_{t}^{(1)}}_{\mbox{$r_2$ terms}}
}_{\mbox{$r$ terms}}),
\label{eqn:D}
\eea
for $t_{\rm min} \le t \le t_{\rm max}$.
For the gauge action,
the symmetrization
and the reduction of the number of lists
can also be applied in this case in a similar way
to Appendix \ref{sec:symmetrization} and \ref{sec:reduction}.

\section{Partial symmetrization of the vertex
at the python level}
\label{sec:symmetrization}
For the gauge action (traced and real),
the summation over all possible permutation in
eq.(\ref{eqn:permutationsum})
can be done partially at the python level, that is,
by creating new lists corresponding to a symmetrized vertex.
It is maybe worth starting from
the vanishing background field~\cite{Luscher:1985wf,Hart:2004bd}.
The color factor in this case (${\cal E}=0$, $\phi_a=\phi^{\prime}_a=0$) is
given by
\be
{\cal C}_{{\rm G}}^{a_1\cdots a_r}
=
{\rm tr}
[I^{a_1} \cdots I^{a_r}+(-1)^rI^{a_r} \cdots I^{a_1} ].
\ee
This is an eigenstate of the inversion and cyclic permutations.
\bea
\sigma\cdot {\cal C}_{{\rm G}}
=
\chi_r(\sigma) {\cal C}_{{\rm G}},
\eea
where the eigenvalues are given by
\be
\chi_r(\sigma)
=
\left\{
\begin{array}{cl}
1,      & \mbox{ for cyclic permutations,} \\
(-1)^r, & \mbox{ for inversion permutation.}
\end{array}
\right.
\ee
By using these properties,
the symmetrized vertex can be written as\footnote{
Here we show the time-momentum space representation,
but the following discussion is also valid for the
coordinate space case.}
\be
V^{a_1 \cdots a_r}_{\mu_1 \cdots \mu_r}
({\bf p}_1,\cdots,{\bf p}_r;t_1,\cdots,t_r)
=
\frac{1}{r!}
\sum_{\sigma\in{\cal S}_r/{\cal Z}_r}
\sigma \cdot {\cal C}_{{\rm G}}^{a_1\cdots a_r}
\sigma \cdot Y^{\prime}_{\mu_1 \cdots \mu_r}
({\bf p}_1,\cdots,{\bf p}_r;t_1,\cdots,t_r),
\ee
with
\be
Y^{\prime}_{\mu_1 \cdots \mu_r}
({\bf p}_1,\cdots,{\bf p}_r;t_1,\cdots,t_r)
=
\sum_{\sigma\in{\cal Z}_r}
\chi_r(\sigma) \sigma \cdot Y_{\mu_1 \cdots \mu_r}
({\bf p}_1,\cdots,{\bf p}_r;t_1,\cdots,t_r).
\label{eqn:Yprimezero}
\ee
In Ref.~\cite{Hart:2004bd},
it is described how to perform the summation in
eq.(\ref{eqn:Yprimezero}) at the python level.
We have seen the nice properties of the color factor of the gauge action
in the case of a vanishing background field.
At first sight, however,
it is nontrivial that the color factor for SF has such properties.
We will discuss this issue in the following subsections,
and explain how to obtain a set for the
partially symmetrized vertex.

\subsection{The inversion permutation for the color factor}
\label{sec:inversionpermuation}
First, let us see the inversion permutation for the color factor
of the gauge action in the SF.
An inverted color factor is written by
(remember that $A=0$ for the gauge action)
\bea
\sigma_{\rm inv}\cdot {\cal C}_{{\rm G}}
^{a_1a_2\cdots a_{r-1} a_r}(x_4,0,B,C,D)
&=&
{\cal C}_{{\rm G}}^{a_ra_{r-1}\cdots a_2a_1}
(x_4,0,B,C,D)
\non\\
&=&
(-1)^r
{\cal C}_{{\rm G}}^{a_1a_2\cdots a_{r-1}a_r}
(x_4,0,\tilde{B},\tilde{C},\tilde{D}),
\label{eqn:inverseC}
\eea
with
\bea
\tilde B
&=&
- B,
\\
(\tilde C_1, \tilde C_2, \cdots,\tilde C_{r-1}, \tilde C_r)
&=&
(C_r,C_{r-1},\cdots,C_2,C_1),
\\
(\tilde D_1, \tilde D_2, \cdots,\tilde D_{r-1}, \tilde D_r)
&=&
(D_r,D_{r-1},\cdots,D_2,D_1).
\eea
Under the inversion permutation, the color factor is not
an eigenstate anymore.
However the structure is still stable but
with different components $\tilde{B}$, $\tilde{C}$ and $\tilde{D}$.

\subsection{Cyclic permutations for the color factor}
\label{subsec:cyclicpermutation}
Secondly, let us consider the cyclic permutation.
A cyclically permuted 
color factor is written as
\bea
\sigma_{\rm cyc}\cdot {\cal C}_{{\rm G}}
^{a_1a_2\cdots a_{r-1}a_r}(x_4,0,B,C,D)
&=&
{\cal C}_{{\rm G}}^{a_2a_3\cdots a_ra_1}
(x_4,0,B,C,D)
\non\\
&=&
{\cal C}_{{\rm G}}^{a_1a_2\cdots a_{r-1}a_r}
(x_4,0,\hat{B},\hat{C},\hat{D}),
\eea
where we have used eq.(\ref{eqn:EIE}).
The explicit forms of the ``hat'' objects are as follows,
\bea
\hat B
&=&
B,
\\
(\hat C_1, \hat C_2, \cdots,\hat C_{r-1}, \hat C_r)
&=&
(C_r-2B,C_1,C_2,\cdots,C_{r-1}),
\\
(\hat D_1, \hat D_2, \cdots,\hat D_{r-1}, \hat D_r)
&=&
(D_r,D_1,D_2,\cdots,D_{r-1}).
\eea

Composite cyclic permutations can be done
in a similar way.
For $1\le j \le r-1$,
\bea
(\sigma_{\rm cyc})^{j}\cdot
{\cal C}_{{\rm G}}^{a_1\cdots a_r}(x_4,0,B,C,D)
&=&
{\cal C}_{{\rm G}}^{a_{j+1}\cdots a_{r-1}a_ra_1a_2\cdots a_j}
(x_4,0,B,C,D)
\non\\
&=&
{\cal C}_{{\rm G}}^{a_1a_2\cdots a_{r-1}a_r}(x_4,0,\hat B,\hat C,\hat D),
\eea
with
\bea
\hat B
&=&
B,
\\
(\hat C_1, \hat C_2, \cdots, \hat C_r)
&=&
(C_{r-j+1}-2B,C_{r-j+2}-2B,\cdots,C_r-2B,C_1,C_2,\cdots,C_{r-j}),
\\
(\hat D_1, \hat D_2, \cdots, \hat D_r)
&=&
(D_{r-j+1},D_{r-j+2},\cdots,D_r,D_1,D_2,\cdots,D_{r-j}).
\eea
As a further example,
we show an opposite cyclic permutation.
\bea
(\sigma_{\rm cyc})^{-1}\cdot
{\cal C}_{{\rm G}}^{a_1a_2\cdots a_{r-1}a_r}(x_4,0,B,C,D)
&=&
{\cal C}_{{\rm G}}^{a_ra_1\cdots a_{r-2}a_{r-1}}(x_4,0,B,C,D)
\non\\
&=&
{\cal C}_{{\rm G}}^{a_1a_2\cdots a_{r-1}a_r}(x_4,0,\hat B,\hat C,\hat D),
\eea
with
\bea
\hat B
&=&
B,
\\
(\hat C_1, \hat C_2, \cdots,\hat C_{r-1}, \hat C_r)
&=&
(C_2,C_3,\cdots,C_r,C_1+2B),
\\
(\hat D_1, \hat D_2, \cdots,\hat D_{r-1}, \hat D_r)
&=&
(D_2,D_3,\cdots,D_r,D_1).
\eea
If one uses eq.(\ref{eqn:translationCD}),
it is not so difficult to see that
\be
(\sigma_{\rm cyc})^{-1}\cdot {\cal C}_{{\rm G}}
=
(\sigma_{\rm cyc})^{r-1}\cdot {\cal C}_{{\rm G}}.
\ee

\subsection{The partially symmetrized vertex for SF}
\label{subsec:SymmetrizedVertexSF}
Let us see how to construct the set for the
partially symmetrized vertex for the SF,
\be
Y^{\prime a_1 \cdots a_r}_{\mu_1 \cdots \mu_r}
({\bf p}_1,\cdots,{\bf p}_r;t_1,\cdots,t_r)
=
\sum_{\sigma\in{\cal Z}_r}
\sigma \cdot Y^{a_1 \cdots a_r}_{\mu_1 \cdots \mu_r}
({\bf p}_1,\cdots,{\bf p}_r;t_1,\cdots,t_r).
\ee
Starting from lists for the unsymmetrized vertex $L^{(r)}[i]$,
one can create corresponding lists in the following way,
\be
\begin{CD}
    @.   L^{(r)}[i]     @>cyclic>>   (\sigma_{\rm cyc})^{j}\cdot L^{(r)}[i]
\mbox{\hspace{15mm}}(1 \le j\le r-1) \\
@.    @VVinverseV          @.                             \\
    @. \sigma_{\rm inv}\cdot L^{(r)}[i]  @>cyclic>> (\sigma_{\rm cyc})^{j}\cdot\sigma_{\rm inv} \cdot L^{(r)}[i]
\mbox{\hspace{7mm}}(1 \le j\le r-1).
\end{CD}
\ee
Note that in the SF case the cyclic and inversion permutations
are not commutative.
An algorithm to obtain an inverted list, $\sigma_{\rm inv}\cdot L^{(r)}[i]$,
from the original list $L^{(r)}[i]$
\be
L^{(r)}[i]
\longrightarrow
\sigma_{\rm inv}\cdot L^{(r)}[i],
\label{eqn:invertedobject}
\ee
is as follows,
\bea
(\mu_1,\mu_2,\cdots,\mu_{r-1},\mu_r)
&\longrightarrow&
(\mu_r,\mu_{r-1},\cdots,\mu_2,\mu_1),
\\
(t_1,t_2,\cdots,t_{r-1},t_r)
&\longrightarrow&
(t_r,t_{r-1},\cdots,t_2,t_1),
\\
({\bf v}_1,{\bf v}_2,\cdots,{\bf v}_{r-1},{\bf v}_r)
&\longrightarrow&
({\bf v}_r,{\bf v}_{r-1},\cdots,{\bf v}_2,{\bf v}_1),
\\
B
&\longrightarrow&
-B,
\\
(C_1,C_2,\cdots,C_{r-1},C_r)
&\longrightarrow&
(C_r,C_{r-1},\cdots,C_2,C_1),
\\
(D_1,D_2,\cdots,D_{r-1},D_r)
&\longrightarrow&
(D_r,D_{r-1},\cdots,D_2,D_1),
\\
f
&\longrightarrow&
(-1)^r f.
\eea
For a cyclically permuted list
$\sigma_{\rm cyc}\cdot L^{(r)}[i]$,
\be
L^{(r)}[i]
\longrightarrow
\sigma_{\rm cyc}\cdot L^{(r)}[i],
\label{eqn:cyclicobject}
\ee
it is given by,
\bea
(\mu_1,\mu_2,\cdots,\mu_{r-1},\mu_r)
&\longrightarrow&
(\mu_2,\mu_3,\cdots,\mu_r,\mu_1),
\\
(t_1,t_2,\cdots,t_{r-1},t_r)
&\longrightarrow&
(t_2,t_3,\cdots,t_r,t_1),
\\
({\bf v}_1,{\bf v}_2,\cdots,{\bf v}_{r-1},{\bf v}_r)
&\longrightarrow&
({\bf v}_2,{\bf v}_3,\cdots,{\bf v}_r,{\bf v}_1),
\\
B
&\longrightarrow&
B,
\\
(C_1,C_2,\cdots,C_{r-1},C_r)
&\longrightarrow&
(C_2,C_3,\cdots,C_r,C_1+2B),
\\
(D_1,D_2,\cdots,D_{r-1},D_r)
&\longrightarrow&
(D_2,D_3,\cdots,D_r,D_1),
\\
f
&\longrightarrow&
f.
\eea
By gathering all lists,
$\{ L^{(r)}[i] \}$,
$\{ \sigma_{\rm inv}\cdot L^{(r)}[i] \}$,
$\{ (\sigma_{\rm cyc})^{j}\cdot L^{(r)}[i] | 1 \le j\le r-1\}$ and
$\{ (\sigma_{\rm cyc})^{j}\cdot \sigma_{\rm inv}\cdot L^{(r)}[i] | 1 \le j\le r-1\}$,
the set $S$ for the partially symmetrized vertex can be composed,
\bea
S&=&
\{ L^{(r)}[i],
\sigma_{\rm cyc}\cdot L^{(r)}[i],
\cdots,
(\sigma_{\rm cyc})^{r-1}\cdot L^{(r)}[i],
\non\\
&&
\sigma_{\rm inv}\cdot L^{(r)}[i],
\sigma_{\rm cyc}\cdot \sigma_{\rm inv}\cdot L^{(r)}[i],
\cdots,
(\sigma_{\rm cyc})^{r-1}\cdot \sigma_{\rm inv}\cdot L^{(r)}[i] \}.
\label{eqn:partiallysymmetrizedfield}
\eea

\section{Reduction of the number of lists}
\label{sec:reduction}
One can reduce the number of lists
by making use of the properties of the background field.
From eq.(\ref{eqn:VIV}) and (\ref{eqn:EIE}),
we found that the color factor obeys the relation
\bea
\lefteqn{{\cal C}_{{\rm G}}^{a_1\cdots a_r}
(x_4,A,B,(C_1,\cdots,C_r),(D_1,\cdots,D_r))}
\non\\
&=&
{\cal C}_{{\rm G}}^{a_1\cdots a_r}
(x_4,A,B,(C_1+\alpha,\cdots,C_r+\alpha),(D_1+\beta,\cdots,D_r+\beta)),
\label{eqn:translationCD}
\eea
for arbitrary real numbers $\alpha,\beta$.
This equation tells us that
one can shift all elements of the $C=(C_1,\cdots,C_r)$
and $D=(D_1,\cdots,D_r)$ to a standard form
such that $C_1\rightarrow 0$ and $D_1\rightarrow0$,
that is, choosing $\alpha=-C_1$ and $\beta=-D_1$.
After performing the shift for $C$ and $D$ of all lists in a set,
there might be some lists
(e.g. two lists $L^{(r)}[i]$ and $L^{(r)}[j]$
with $i\neq j$)
which have an identical configuration $C[i]=C[j]$ and $D[i]=D[j]$
(of course the Lorentz index configuration,
the coordinate configuration\footnote{
In the momentum space representation,
the replacement $x[i] \rightarrow (t[i],{\bf v}[i])$
is understood.}, $A$ and $B$ in
the two lists should be identical beforehand,
but their amplitudes do not have to be so).
In this case, we can merge the two lists into a list
by summing the amplitude factor $f[i]+f[j]$.
In this way one can reduce the number of lists.
Actually this achieves a few percents reduction of the size of a set
for typical gauge actions.

\section{$\eta$ derivative of the vertex}
\label{sec:etaderivative}
When one calculates the SF coupling,
the $\eta$ derivatives of the vertices are required.
We can easily obtain them from the same python output for the
usual (non $\eta$-derivative) vertices.
Since the information of the background field
($V(x_4)$, ${\cal E}$, $\phi_a^{\prime}$ and $\phi_a(x_4)$
which have $\eta$ dependence)
is encoded in the color factor,
we need to consider the color factor only for our purpose.
For example, the explicit form of
the $\eta$ derivative of the color factor
for the gauge action ($A=0$) is given by
\bea
\lefteqn{
\frac{\partial {\cal C}_{{\rm G}}^{a_1\cdots a_r}(x_4,0,B,C,D)}{\partial \eta}}
\non\\
&=&
\left(\frac{2\sqrt{3}a^2B}{LT}\right)
{\rm tr}
\left[
I^{a_1} \cdots I^{a_r} I^8
e^{ ia^2{\cal E}B}
-(-1)^r
I^8 I^{a_r} \cdots I^{a_1}
e^{-ia^2{\cal E}B}
\right]
e^{\frac{i}{2}\sum_{u=1}^r
(C_u\phi_{a_{u}}^{\prime}
+D_u\phi_{a_{u}}(x_4))}
\non\\
&+&
\frac{i}{2}\sum_{u=1}^r
\left(
 C_u\pal_{\eta}\phi_{a_{u}}^{\prime}
+D_u\pal_{\eta}\phi_{a_{u}}(x_4)
\right)
{\cal C}_{{\rm G}}^{a_1\cdots a_r}(x_4,0,B,C,D),
\label{eqn:etacolor}
\eea
where we have used eq.(\ref{eqn:etaE}).
The information of $B$, $C$ and $D$ is enough to identify the above term.
When writing down the $\eta$ derivative vertex
on a diagram calculation code,
one needs additional phases $\pal_{\eta}\phi_{a}^{\prime}$
and $\pal_{\eta}\phi_{a}(x_4)$ which are
given in Table \ref{tab:phis} in analytic forms,
and an additional color matrix
\be
{\rm tr}
\left[
I^{a_1} \cdots I^{a_r} I^8
e^{ ia^2{\cal E}B}
-(-1)^r
I^8 I^{a_r} \cdots I^{a_1}
e^{-ia^2{\cal E}B}
\right].
\ee
This has a form similar to that in eq.(\ref{eqn:colormatrixG})
of the usual vertex
apart from the presence of $I^8$ and the relative sign between the two terms.

\end{appendix}

\providecommand{\href}[2]{#2}\begingroup\raggedright\endgroup

\end{document}